\documentclass[twocolumn,pra,nofootinbib]{revtex4-2}

\usepackage{amssymb,amsmath,mathtools,bm}
\usepackage[colorlinks]{hyperref}
\usepackage{color}
\usepackage{graphicx}

\begin{document}
\title{\bf Fluctuations of atomic energy levels due to axion dark matter}
\author{V. V. Flambaum}\email{v.flambaum@unsw.edu.au}
\author{I. B. Samsonov}\email{igor.samsonov@unsw.edu.au}
\address{School of Physics, University of New South Wales,
Sydney 2052, Australia}

\begin{abstract}
The amplitude of the pseudoscalar (axion) or scalar field fluctuates on a time scale of order of  million field oscillation periods which is a typical coherence time in the virialized axion galactic dark matter halo model. This causes fluctuations of frequencies of atomic clocks on the same time scale. We show that this effect may be employed to search for the axion and scalar field dark matter with atomic and nuclear clocks. We re-purpose the results of the atomic clocks experiments comparing the variations of frequencies of hyperfine transitions in Rb and Cs atoms as well as in hydrogen atom vs cavity frequency fluctuations, and extract new limits on the axion coupling constant $f_a$ for masses in the range $2.4\times 10^{-17}\text{ eV}\lesssim m \lesssim 10^{-13}\text{ eV}$. We also show that similar energy shifts arise in the second-order perturbation theory with linear in the pseudoscalar field interaction. These shifts may be potentially measured with nuclear clocks based on the low-energy transition in $^{229}$Th nucleus. We propose a procedure which could, in principle, help determine the axion mass if the axion dark matter signal is present in experimental data sets.
 \end{abstract}

\maketitle

\section{Introduction}
Scalar and pseudoscalar particles represent promising dark matter candidates which can fully saturate the local dark matter density $\rho_\text{DM}\approx 0.4\text{ GeV/}\text{cm}^3$ \cite{Preskill,Abbott,Dine}. If the mass of each of these particles is low, $m\ll 1\text{ eV}$, their number per De Broglie wavelength must be large, and the ensemble of these particles may be considered as a classical field oscillating harmonically in every particular point of space, $\phi=\phi_0 \cos (\omega t) $. The oscillation frequency $\omega$ is approximately equal to the dark matter particle mass $m$, $\omega \approx m$, since the kinetic energy is small, $E_k \sim 10^{-6} m$ (in this paper, we assume that the dark matter particles are virialized within the standard dark matter halo model, see, e.g., Refs.~\cite{DMhalo,halo1,DMhalo2}). Interaction of the standard  model particles (electron, photon, quarks, gluons) with this dark matter field  produces oscillating shifts of atomic energy levels which have been searched for in a number of experiments, see, e.g., Refs.~ \cite{Arvanitaki,Stadnik,Stadnik2,DyCs,RbCs,YbCs,HSi,Tretiak,Yb,Huntemann,DyQuartz,NPL}. 

A general problem is that the mass of dark matter particle is unknown; therefore, one should do Fourier  analysis of the data to separate the oscillating signal. 
However, if the interaction is proportional to the scalar field squared, 
\begin{equation} 
\label{1}
V = -g_f M_f \phi^2 \bar \psi \psi -\frac{g_{\gamma}}{4} \phi^2 F_{\mu \nu}  F^{\mu \nu}+\ldots \,, 
\end{equation}
the energy shift has a non-oscillating contribution owing to the identity
\begin{equation}
\label{2}
    \phi^2 =\frac12\phi_0^2 [1+  \cos (2 \omega t)]\,.
\end{equation}
In Eq.~(\ref{1}), $\psi$ is a Dirac fermion field with mass $M_f$, $F_{\mu\nu}$ is the Maxwell field strength, $g_f$ and $g_\gamma$ are the corresponding coupling constants. In general, the scalar field may couple to other Standard Model fields, denoted by ellipsis in Eq.~(\ref{1}), which are not important in the present consideration. 

Effects of quadratic-in-$\phi$ interaction in Eq. (\ref{1}) may be described as an apparent  variation of the fine structure constant, $\alpha'=\alpha (1 + g_{\gamma} \phi^2)$, and masses of elementary particles, $M'_f=M_f(1+ g_f  \phi^2)$, see, e.g., Refs.~\cite{Arvanitaki,Stadnik}. For example, the mass shift immediately follows from comparison of interaction with the scalar filed $ -g_f M_f  \phi^2 \bar \psi \psi $ and fermion mass term in the Lagrangian $-M_f \bar \psi \psi$. Dependence of atomic transition frequencies on $\alpha$, quark masses and $\phi^2$ was studied in works \cite{PRLWebb,PRAWebb,CanJPh,Tedesco,Stadnik,Stadnik2,Borschevsky,csquarks}. Atomic spectroscopy methods have already allowed one to improve earlier cosmological limits  on the interaction strength of low mass scalar field $\phi^2$ with photons, electrons and quarks  by 15 orders in magnitude \cite{Stadnik,Stadnik2}. These limits have recently been revisited in Ref.~\cite{BBN-new} due to Big Bang Nucleosynthesis considerations. The experimental results were obtained by the measurements of oscillating frequency ratios of electron transitions in Dy/Cs \cite{DyCs}, Rb/Cs \cite{RbCs}, Yb/Cs \cite{YbCs}, Sr/H/Si cavity \cite{HSi}, Cs/cavity \cite{Tretiak},  Yb/Yb/Sr \cite{Yb,Huntemann}, Rb/Quartz \cite{DyQuartz} where  effects of the variation of frequencies may be interpreted as variation of $\alpha$ and fermion masses. In Ref.~\cite{GNOMEdE} it was proposed to search for the scalar field dark matter with interaction (\ref{1}) by measuring fluctuations of the scalar field amplitude using magnetometer and optical atomic clock networks. In the case of linear-in-$\phi$ interaction, the use of a network of precision-measurement tools for searches of wave-like dark matter was proposed in Ref.~\cite{Devevianko2018}.

The rest of this paper is organized as follows. In Sec.~\ref{SecI} we consider quadratic-in-$\phi$ atomic energy level corrections arising in the first order of perturbation theory and demonstrate that the mean value of these shifts should be (approximately) equal to the standard deviation due to the stochastic nature of the axion field amplitude. This relation allows us to find new lab-based limits on the axion decay constant $f_a$ by re-purposing the results of the experiments \cite{RbCs} and \cite{HSi}, see Sec.~\ref{SecLimits}. Then, in Sec.~\ref{SecSignal} we propose a procedure which, in principle, would allow one to identify the axion dark matter signal in experiments measuring fluctuations of energy level shifts with atomic clocks. This procedure utilizes the fact that the atomic energy levels fluctuations caused by axion dark matter should have a different statistical distribution from ordinary noise in the detector. In Sec.~\ref{Sec2Order} we show that quadratic-in-$\phi$ contributions to the atomic energy level shifts appear also in the second order of perturbation theory. We compare these contributions to the corresponding first-order energy level corrections by considering the example of low-lying isomeric state in $^{229}$Th nucleus. Section \ref{SecSummary} is devoted to a summary and discussion of the results of this paper.

We use natural units with $\hbar=c=1$.


\section{First-order perturbation theory energy level corrections due to quadratic axion-nucleon interaction}
\label{SecI}

In this section, we focus on the quadratic in the QCD axion field $\phi$ interaction with a nucleon as in Eq.~(\ref{1}). In Ref.~\cite{KimPerez} it was shown that this interaction originates from the standard QCD $\theta$-term
\begin{equation}
\label{thetaQCD}
    \frac{g^2\theta}{32\pi^2}\tilde G^{l\,\mu\nu}G^l_{\mu\nu}\,,
\end{equation}
with $\theta=\phi/f_a$, $f_a$ is the axion decay constant, $g$ is the strong interaction coupling constant, $G^l_{\mu\nu}$ is the gluon field strength and $\tilde G^{l\,\mu\nu}$ is its Hodge dual. Indeed, this axion-gluon interaction implies a variation of the pion mass \cite{Ubaldi}, 
\begin{equation}
\label{deltaPi}
\frac{\delta m_{\pi}}{m_{\pi}} \approx  -  0.05 \theta^2\,.
\end{equation}
This causes the corresponding variations of nuclear magnetic moment, nuclear mass and radius since these quantities depend on the pion mass, see Refs.~\cite{Thomas,Tedesco,Wiringa1,Wiringa2,Dinh}. As a result, the atomic energy level shift is proportional to the scalar field squared,
\begin{equation}
\label{E1}
    E \propto \phi^2\,.
\end{equation}

We will study statistical properties of such energy shifts within two models of the ultralight axion dark matter. In the first one, the axion field at the observation point is represented by a monochromatic wave with a uniformly distributed random phase and random amplitude with Rayleigh distributions, while in the second one we consider the axion field as a wave packet appearing due to a spread of velocities of dark matter particles. Although the first model may be considered as a simplified version of the second one, it allows for a simpler treatment and exact analytical results. 

\subsection{Monochromatic wave model}
\label{SecMonochrom}

The axion field dark matter may be modeled by a monochromatic scalar wave with angular frequency $\omega = m(1+v^2/2)$, where $v\sim 10^{-3}$ is the most probable speed of dark matter particles in the standard dark matter halo model \cite{DMhalo,halo1,DMhalo2},
\begin{equation}
\label{SingleWave}
    \phi = \phi_0 \cos(\omega t + \varphi)\,.
\end{equation}
Here $\varphi\in[0,2\pi)$ is a uniformly distributed random phase and $\phi_0$ is a random amplitude governed  by the Rayleigh distribution \cite{Derevianko}
\begin{equation}
p(\phi_0)=
\frac{2 \phi_0}{\phi_\text{DM}^2} \exp{\left(-\frac{\phi_0^2}{\phi_\text{DM}^2}\right)}\,, 
\label{distribution}
\end{equation}
where $\phi_\text{DM} = \sqrt{2\rho_\text{DM}}/m$ is an average amplitude of the dark matter field. The variance of the scalar field amplitude stems from stochastic nature of phases of dark matter particles near the Solar system but the spread of speeds of these particles is ignored. This spread of speeds of dark matter particles will be accounted for in the next subsection. 

The model (\ref{SingleWave}) ignores also possible nonvirialized dark matter streams \cite{streams} and composite dark matter structures such as boson stars \cite{BosonStars} or topological defects \cite{TopDefects}, which are to be studied separately.

Substituting Eq.~(\ref{SingleWave}) into (\ref{E1}), we have the following axion field dependence of the atomic energy shift
\begin{equation}
\label{Energy2}
    E \propto \frac12\phi_0^2[(1+\cos(2\omega t+2\varphi)]\,.
\end{equation}
Let us assume that the experimental integration time $t_1$ significantly exceeds the oscillation period $T= \pi/\omega\approx\pi/m$, $t_1\gg T$. In this case, the oscillating term in Eq.~(\ref{Energy2}) averages to zero,  
\begin{equation}
\begin{aligned}
\overline{\cos (2 \omega t+2\varphi)}&=
\frac1{t_1}\int_0^{t_1} \cos(2\omega t+2\varphi)dt \\&
= \frac{\sin(2\omega t_1+2\varphi)}{2\omega t_1} \to 0
\end{aligned}
\end{equation}
for $\omega t_1\gg1$. As a result, the quadratic interaction (\ref{1}) implies the following shift of an atomic energy level in the first order of perturbation theory
\begin{equation}
E \equiv \langle V \rangle = C  (\phi_0)^2, 
\label{EnergyShift}
\end{equation}
where $C$  is a time-independent constant. 

The problem is that only the time dependence of atomic energy level shifts produced by new interactions  can be measured accurately. Usually, the time-independent contribution to the energy shift is hidden by uncertainties of theoretical values of energies in multielectron atoms. 

In the case of quadratic interaction (\ref{1}) with the scalar field (\ref{2}) this problem may be addressed as follows. The amplitude of the scalar or pseudoscalar (axion) field $\phi_0$ fluctuates on the time scale $\tau \sim 10^6 T$  (coherence time), see, e.g., Ref.~\cite{Derevianko}.  This causes fluctuations of the energy shift (\ref{EnergyShift}) of atomic, molecular and nuclear transition energies. One can set the integration time $t_1$ much smaller than the coherence time $\tau$ but much larger than the oscillation period $T$,
\begin{equation}
\label{t-condition}
T\ll t_1 \ll \tau\,.
\end{equation}
Repeating these measurements $N$ times such that the total measurement time $t = N t_1$ exceeds the coherence time $\tau$, 
\begin{equation}
t \gg \tau \,,
\label{t-condition2}
\end{equation}
one can measure variance of fluctuations of the energy shift $E$ 
\begin{equation}
\sigma_\phi^2=\overline{(E - \overline{E})^2}=\overline{(E^2)} -  (\overline{E})^2 = (\overline{E})^2\,.
\label{deltaE}
\end{equation}
Here we used the probability distribution function (\ref{distribution}) for the scalar field amplitude.
As a result, Eq.~(\ref{deltaE}) allows us to identify a theoretically calculated atomic energy level shift with experimentally measured standard deviation of this shift,
\begin{equation}
\label{sigmaE}
    \sigma_\phi = \bar E\,.
\end{equation}

Note that the atomic energy level shift (\ref{EnergyShift}) due to quadratic in the axion field interaction (\ref{1}) is a random variable with exponential distribution 
\begin{equation}
    p_\phi(E) = \left\{ 
    \begin{array}{ll} 
        \frac1{\bar E}e^{-E/\bar E}\quad &\text{for }E\geq 0\,,\\
        0& \text{for }E<0\,,
    \end{array}
    \right.
\end{equation}
since the axion field amplitude follows the Rayleigh distribution. The relation (\ref{sigmaE}) is simply a property of the exponential distribution. Another important feature of this distribution is the presence of non-vanishing higher statistical moments such as skewness $S$ and kurtosis $K$:
\begin{equation}
\label{SK}
    S\equiv \frac{\overline{( E - \bar E)^3}}{\sigma^3} = 2\,,\qquad
    K\equiv \frac{\overline{( E - \bar E)^4}}{\sigma^4} = 9\,.
\end{equation}
Thus, the axion signal should manifest itself in the experimental data through properties of the exponential distribution.

Equation (\ref{sigmaE}) is derived within the assumption that the energy shift fluctuations are caused solely by the axion field dark matter. In reality, one has to add the effect of the noise in the detector. Let $p_\text{noise}(E)$ be a probability distribution function of this noise with standard deviation $\sigma_\text{noise}$ and vanishing mean. Then the experimentally measured atomic energy level shift is a random variable which follows a convolution of these two distributions,
\begin{equation}
    p(E) = \int_{-\infty}^{\infty} p_\phi(E-E')p_\text{noise}(E')dE'\,.
\label{convolution}
\end{equation}
As a result, the standard deviation for this combined distribution is greater than the one for exponential distribution (\ref{sigmaE}),
\begin{equation}
\label{sigmaLimit}
    \sigma > \bar E\,.
\end{equation}
The values of the higher statistical moments (\ref{SK}) are also different for this distribution. They will be considered in more detail is Sec.~\ref{SecSignal} where a simulation of experimental data set will be studied.

The relation (\ref{sigmaLimit}) will be used in the next section for extracting limits on the axion parameter space from experiments measuring energy level shifts with atomic clocks. This method is broadband as it does not require performing Fourier analysis of the data. The principal assumption in this approach is that the measurement time $t_1$ satisfies the conditions (\ref{t-condition}) and (\ref{t-condition2}). Given that the energy shift  oscillation period $T=\pi/m$, we can convert these conditions to the constraints on the scalar field mass if the time intervals $t_1$ and $t$ are experimentally fixed,
\begin{equation}
\label{conditions}
    \max \{ \pi/t_1 , 10^6\pi/t \} \ll m \ll 10^6 \pi /t_1\,.
\end{equation}
Thus, an experiment measuring fluctuations of frequencies of atomic clocks during the time $t$ is suitable for searches of the axion and scalar dark matter with particle mass in the range (\ref{conditions}).

This approach may be efficient when axion or scalar mass is not too small. For instance, assume that the averaging time is $t_1 > 10^{-6}$ s, then the coherence time obeys $\tau > t_1> 10^{-6}$ s, and the oscillation period is $T \sim 10^{-6} \tau > 10^{-12}$ s. Assuming also that the total measurement time is about one day, we have $\tau <t \sim 10^5$~s and obtain the range of dark matter particle masses  $ 10^{-13}$ eV $< m <$ 0.01 eV. The QCD axion with the mass $m\sim 10^{-5}$~eV falls within this region.

Note that a similar proposal of exploring fluctuations of the dark matter amplitude has been recently presented in Ref. \cite{GNOMEdE}. The novel feature of the present work is the idea of using experimental value of the variance and higher moments to find the limits on the axion decay constant. This idea will be illustrated by the following two examples.

\subsection{Wave packet model}
\label{SecPacket}

Equation (\ref{SingleWave}) may be considered as a toy model for the ultralight axion dark matter because it does not take into account a spread of frequencies in this field due to stochastic nature of velocities of dark matter particles in Galaxy. A more realistic axion (or scalar) field dark matter model was developed in Ref.~\cite{Derevianko} where this field in a point of observations is represented by a wave packet
\begin{equation}
    \phi(t)=\frac{\sqrt{\rho_\text{DM}}}{m}
     \sum_{i=1}^n \alpha_i \sqrt{f(v_i)\Delta v} \cos\left[ m\left(1+\frac{v_i^2}{2}\right)t + \varphi_i \right],
\label{phiSum}
\end{equation}
where the sum is taken over the speeds $v_i$ of dark matter particles with the distribution from the standard dark matter halo model \cite{DMhalo,halo1,DMhalo2}
\begin{equation}
\label{fSHM}
    f(v) = \frac{v}{\sqrt{\pi}v_0 v_\text{obs}}e^{-(v+v_\text{obs})^2/v_0^2}
    (e^{4v v_\text{obs}/v_0^2}-1)
\end{equation}
with $v_0\approx 220$ km/s the speed of the local rotation curve and $v_\text{obs} \approx 232$ km/s the speed of the Sun in the galactic rest frame.\footnote{The function (\ref{fSHM}) is normalized as $\int_0^\infty f(v)dv =1$. In reality, the dark matter particle velocities in the local rotation curve vary from the escape velocity in the Solar system $v_\text{min}=72$ km/s to the galactic escape velocity $v_\text{max}\approx600$ km/s. Restricting the velocities to this interval slightly changes the overall coefficient in (\ref{fSHM}), but we ignore this effect for simplicity.} 

The sum in Eq.~(\ref{phiSum}) should have sufficiently large number of terms $n$ such that the variations of speeds of particles in each interval $[v_i,v_i+\Delta v]$ may be neglected, and these particles may be described by a monochromatic wave with uniformly distributed random phase $\varphi_i$ and amplitude proportional to a random variable $\alpha_i$ with Rayleigh distribution
\begin{equation}
    p(\alpha_i) = \alpha_i e^{-\alpha_i^2/2}\,.
\end{equation}

Recall that we consider experiments measuring atomic or molecular energy level shifts which are quadratic with respect to the axion field, see Eq.~(\ref{E1}). Assume that such an experiment repeatedly measures atomic energy level shifts at time instances $t_i$ such that $\Delta t=t_{i+1}-t_i=t_1$ obeys the conditions (\ref{t-condition}). Therefore, fast oscillations of the axion field are averaged out during the integration time,
\begin{align}
    &\langle \phi^2 \rangle \equiv \frac1{\Delta t} \int_{t_i}^{t_{i+1}} \phi^2(t)dt \nonumber\\
    &=\frac{\rho_\text{DM}}{2m^2} \sum_{j,k=1}^n \alpha_j \alpha_k
    \Delta v\sqrt{f(v_j) f(v_k)} \label{22} \\
    &\times \frac{\sin[m\Delta t(v_k^2-v_j^2)/4] \cos[m\bar t_i (v_k^2-v_j^2)/2 +\Delta\varphi_{kj}] }{m\Delta t(v_k^2-v_j^2)/4}\,,
    \nonumber
\end{align}
where $\Delta\varphi_{kj} = \varphi_k-\varphi_j$ and $\bar t_i = (t_i+t_{i+1})/2 = (i+\frac12)\Delta t$. 

Assume now that the integration time $\Delta t$ is such that the following conditions 
\begin{equation}
    m\Delta t(v_k^2-v_j^2)/4 \ll1 \quad \forall k,j\leq n
\end{equation}
are satisfied. In this case Eq.~(\ref{22}) reduces to
\begin{equation}
\begin{aligned}
    \langle \phi^2(t_i) \rangle \approx &\frac{\rho_\text{DM}}{2m^2} \sum_{j,k=1}^n \alpha_j \alpha_k
    \Delta v\sqrt{f(v_j) f(v_k)}
    \\&\times\cos[m t_i (v_k^2-v_j^2)/2 +\Delta\varphi_{kj}]\,.
\end{aligned}
\label{24}
\end{equation}
This equation specifies time fluctuations of atomic energy level shifts due to ultralight axion dark matter,
\begin{equation}
    E_i \equiv E(t_i) = C \langle \phi^2(t_i) \rangle\,,
\label{25}
\end{equation}
where $C$ is a time-independent coefficient. Equations (\ref{24}) and (\ref{25}) show that the time fluctuations of the atomic energy levels due to axion dark matter is described by a wave packet with a spread of frequencies due to stochastic nature of speeds of dark matter particles in the point of observation.

Eq.~(\ref{24}) may be conveniently represented as a sum of time-independent and oscillating terms,
\begin{equation}
\begin{aligned}
    \langle \phi^2(t_i) \rangle & =
    \frac{\rho_\text{DM}}{2m^2} \sum_{j=1}^n \alpha_j^2 f(v_j)\Delta v \\&+
    \frac{\rho_\text{DM}}{m^2} \sum_{j>k}^n \alpha_j \alpha_k
    \Delta v\sqrt{f(v_j) f(v_k)}
    \\&\times\cos[m t_i (v_k^2-v_j^2)/2 +\Delta\varphi_{kj}]\,.
\end{aligned}
\label{26}
\end{equation}
Hence, after averaging over a time interval $t=N\Delta t$ significantly exceeding periods of all oscillating terms, we have a non-vanishing mean energy level shift
\begin{equation}
\bar E = \frac1N\sum_{i=1}^N E_i = C  \frac{\rho_\text{DM}}{2m^2} \sum_{j=1}^n \alpha_j^2 f(v_j)\Delta v\,.
\label{27}
\end{equation}

Note that the coherence time in fluctuations of the atomic energy shifts due to axion dark matter is usually defined as $\tau\sim 10^6 \pi/m$. This definition, however, needs to be formalized because these shifts are described by a wave packet (\ref{24}) with no fixed frequency. For this purpose, we consider a time correlation function of relative energy shifts,
\begin{align}
    R(\tilde{t})&\equiv \langle \Delta E(t' +\tilde{t})\Delta E(t') \rangle_{t'}
    \nonumber\\
    &=\frac1{t_w}\int_{0}^{t_w} \Delta E(t'+\tilde{t})\Delta E(t') dt'\,,
\label{correlator}
\end{align}
where $t_w$ is the averaging time window  ($\tau \ll t_w < t -\tilde{t}$), $\Delta E(t') = E(t')-\bar E$, $E(t')$ is given by Eqs.~(\ref{24}) and (\ref{25}), and $\bar E$ is defined in Eq.~(\ref{27}). The coherence time $\tau$ may be defined now as a time interval needed for the correlation function (\ref{correlator}) to fall to half its original value,
\begin{equation}
    R(\tau) = \frac12R(0)\,.
\label{Rtau}
\end{equation}
Given this definition of the coherence time, we stress that the applicability of this approach is given by the conditions (\ref{t-condition}) and (\ref{t-condition2}).

The energy level shifts (\ref{25}) represent a random variable which approximately follows the exponential distributions because Eq.~(\ref{24}) is a weighted sum of a product of Rayleigh distributed random amplitudes $\alpha_i$. Therefore, the equation (\ref{sigmaE}) is now satisfied only approximately,
\begin{equation}
    \sigma_\phi \approx \bar E\,,
    \label{31}
\end{equation}
where $\sigma_\phi$ is the standard deviation of energy level shifts (\ref{25}). Higher statistical moments (\ref{SK}) develop their values also only approximately, $S\approx 2$, $K\approx 9$.

When the noise in the detector is taken into considerations, the standard deviation increases, and the relation (\ref{31}) turns into the constraint (\ref{sigmaLimit}). Thus, this limit can be used for constraining the parameter space in the ultralight axion dark matter model from the experiments measuring atomic energy level shifts with atomic clocks.


\section{Limits from atomic clocks experiments}
\label{SecLimits}

In the previous section, we have shown that the axion signal may be, in principle, detected in the experiments comparing atomic energy level shifts between different atomic clocks and studying their statistical properties. The limits on the axion parameter space may be found even without full data sets of these experiments, but just with known leading statistical moments such as the standard deviation, skewness and kurtosis, because the relations (\ref{sigmaE}) and (\ref{SK}) may be considered as signatures of stochastic fluctuations of the axion field amplitude. 

In a real experiment, the noise in the detector can hide the axion signal and spoil the properties (\ref{sigmaE}) and (\ref{SK}). In Sec.~\ref{SecMonochrom}, we have shown that this noise can increase the value of the standard deviation, and the relation (\ref{sigmaLimit}) should be rather used for extracting limits from experiments measuring atomic energy level shifts with atomic clocks. We stress that the relation (\ref{sigmaLimit}) holds for both models of the axion dark matter considered in the previous section.

The region of applicability of the constraint (\ref{sigmaLimit}) is given by conditions (\ref{conditions}) which originate from relations (\ref{t-condition}) and (\ref{t-condition2}). These relations, in particular, mean that the coherence time should be much larger than the integration time in one measurement. In other words, there are many 
energy shift measurements per axion dark matter coherence time, which, if the detector noise were vanishing, would have nearly the same values. These energy shifts all together would still follow the exponential distribution (at least approximately in the case of wave packet considered in Sec.~\ref{SecPacket}) if the total experimental time $t$ significantly exceeds the coherence time. Indeed, outside the coherence time the values of energy shifts will differ and if they cover the whole range of allowed values, the variance and higher moments do not change because of the repetition of energy shifts within coherence time.  This also immediately follows from the definition of variance (and similar for other moments):
\begin{equation}
\sigma^2=\frac1{N-1}\sum_{i=1}^N (E_i -\overline{E})^2.
\end{equation}
If we multiply each term in this sum by factor $n$, there will be compensating increase of $N$ in denominator, $N'=N n$. 
On the other hand, if the total duration of the experiment $t$ appears shorter  than the coherence time $\tau$, then the axion-induced energy shifts will have approximately equal values with a  small variance and the limit (\ref{sigmaLimit}) cannot be applied any more. Thus, the relation (\ref{sigmaLimit}) is applicable in the region of axion mass (\ref{conditions}).

Below we consider two experiments measuring atomic energy level shifts in Rb/Cs \cite{RbCs} and H/Si cavity \cite{HSi} pairs and extract new limits on the axion coupling constant $f_a$ from experimentally measured standard deviation.

\subsection{Limits from Rb/Cs experiment}

Measurements of time dependence of the ratio of frequencies of Rb and Cs hyperfine transitions were implemented in the work \cite{RbCs}. Using calculations in Refs. \cite{Tedesco,KimPerez}, we find this ratio in the form
  \begin{equation}
\frac{\delta (\nu_\text{Rb}/\nu_\text{Cs})}{ \nu_\text{Rb}/\nu_\text{Cs}}=
10^{-16}\frac{(1+\cos(2 m t))}{m_{15}^2 f_{10}^2}\frac{\phi_0^2}{\phi_\text{DM}^2}\,, 
\end{equation}
where $m_{15}\equiv m/(10^{-15}$ eV), $f_{10}\equiv f_a/(10^{10}$ GeV), $m$  and $f_a$ are  the axion mass and interaction constant. The reported standard deviation in measurements of variation of the ratio of frequencies is $\sigma = 3 \times 10^{-15}$ \cite{RbCs}. The averaging time in this experiment is $t_1=864$ s, and the number of measurements is $N=100814$. Substituting these values into the conditions (\ref{conditions}), we find the limits for the axion mass in the range  $2.4 \times 10^{-17}$ eV $\ll m\ll 2.4 \times 10^{-12}$ eV:
\begin{equation}
f_a > 1.8 \times 10^{9} \text{GeV}  \left(\frac{10^{-15} \text{eV}}{m}\right) \,.
\label{flimit1}
\end{equation}
The corresponding exclusion region is shown in Fig.~\ref{fig:limits} by the blue area. Although this limit is still many orders in magnitude weaker than the QCD axion line, Eq.~(\ref{flimit1}) gives a new constraint on the axion coupling $f_a$ for axion masses in the range $2.4\times 10^{-17}\text{ eV}\lesssim m \lesssim  10^{-13}\text{ eV}$ which is not covered by other lab-based experiments. Note that in presenting these constraints in Fig.~\ref{fig:limits} we assume that the parameters $f_a$ and $m$ are independent while they are related as $f_a m\approx f_\pi m_\pi$ for the canonical QCD axion.

\subsection{Limits from H/Si cavity experiment}
\label{SecHSi}
\begin{figure}
    \centering
    \includegraphics[width=8.5cm]{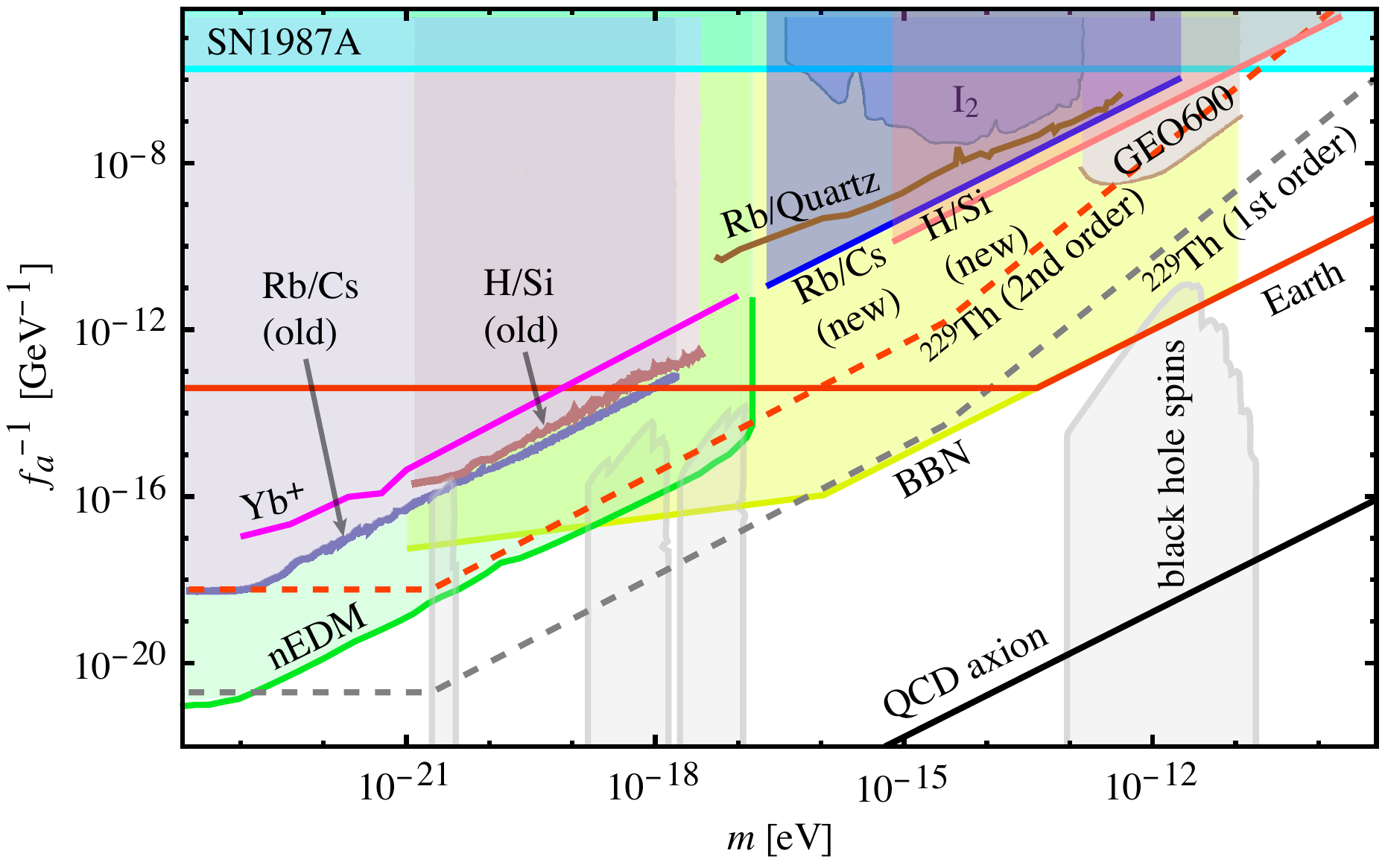}
    \caption{New limits on the axion coupling constant $f_a$ from re-purposing the results of the Rb/Cs atomic clocks experiments \cite{RbCs} (blue shaded region) and H/Si cavity experiment \cite{HSi} (pink shaded region). These limits are compared with the earlier found constraints from the same experiments obtained in Ref.~\cite{KimPerez} (gray shaded regions labeled as Rb/Cs (old) and H/Si (old) respectively). Dashed red line represents projected sensitivity of $^{229}$Th nuclear clocks to the axion field through the second-order energy shift in the perturbation theory, and dashed gray line corresponds to the first-order perturbation theory correction found in Ref.~\cite{KimPerez}. The black straight line in the right bottom corner is the QCD axion benchmark line with $f_a m \approx f_\pi m_\pi$. For comparison, we included also the constraints from I$_2$ molecular spectroscopy experiment \cite{I2}, GEO 600 gravitational detector \cite{GEO600}, nuclear spin precession experiment (nEDM) \cite{nEDM}, big bang nucleosynthesis (BBN) \cite{BBN}, supernova explosions (SN1987A) \cite{SN1987}, oscillation of nuclear charge radius in Yb$^+$ experiment \cite{Yb}, Rb/Quartz oscillator experiment \cite{DyQuartz}, and analysis of black hole spins \cite{spins1,spins2}. The red solid line labeled ``Earth'' represents constraints due to possible axion emissions from the Earth \cite{Earth}.}
    \label{fig:limits}
\end{figure}

Similar limit may be obtained from the comparison of the hydrogen hyperfine transition with the silicon cavity eigenmode performed in Ref.~\cite{HSi}. Dependence of the ratio of corresponding  frequencies on the fundamental constants 
has been obtained in Refs.~\cite{Tedesco,Borschevsky}
  \begin{equation}
\frac{\nu_\text{H}}{\nu_\text{Si}}\propto \alpha^3 R(Z \alpha) \frac{m_e}{m_p}g_p\,,
\end{equation}
where $m_e$ and $m_p$ are electron and proton masses, respectively, $g_p$ is the proton magnetic $g$-factor, $\alpha$ is the fine structure constant, $Z$ is the nuclear charge and $R(Z \alpha)$ is the relativistic factor which for hydrogen and silicon is close to 1.  Using calculations presented in  Ref.~\cite{KimPerez} we obtain 
 \begin{equation}
\frac{\delta (\nu_\text{H}/\nu_\text{Si})}{ \nu_\text{H}/\nu_\text{Si}}=10^{-15}\frac{(1+\cos(2 m t))}{m_{15}^2 f_{10}^2}\frac{\phi_0^2}{\phi_\text{DM}^2}\,.
\end{equation}
Equating this frequency variation to the dispersion of the experimental data in Ref.~\cite{HSi}, $\sigma \approx 3\times 10^{-15}$, we find the limit on the axion decay constant:
\begin{equation}
f_a > 5.8 \times 10^{9}\text{ GeV}\left( \frac{10^{-15}\text{eV}}{m} \right)\,.
\end{equation}
Although this constraint is comparable with that in Eq.~(\ref{flimit1}), it applies to a slightly different axion mass range $7.3\times 10^{-16}\text{ eV}< m < 1.9\times 10^{-10}\text{ eV}$ which corresponds to the integration time $t_1=10.7$~s \footnote{This integration time follows from Ref.~\cite{HSi}: The total measurement time in the H/Si cavity experiment 2826942~s should be divided by 368 data points presented in Fig.~1b in this work, and by a factor 720 which represents the decimation of the original data set. Thus, $t_1= 2826942\text{ s}/(368\times 720)\approx 10.7$~s.}.
The corresponding exclusion region is shown in Fig.~\ref{fig:limits} by a pink area.

This effect may also be measured in molecules where vibrational and rotational transitions are sensitive to variation of nucleon mass, see, e.g., Ref.~\cite{I2,KimPerez}. In fact, variance in the fluctuations of energy levels was measured in numerous  papers searching for the linear drift of the fundamental constants. This variance is linked to statistical error of the drift measurements.
 

\section{Possible signature of the axion signal}
\label{SecSignal}

In Refs.~\cite{Devevianko2018,GNOMEdE} it was shown that the axion field may manifest itself in correlated fluctuations of energy shifts in a network of atomic clocks or magnetometers. Such a correlation is possible if the detectors in a network are separated by a distance not exceeding the dark matter particle correlation lengths. Here we will demonstrate that, in principle, it is possible to find a signature of the ultralight axion dark matter even with a single atomic clock measuring relative atomic energy level shift within a sufficiently long period of time. The main idea is that fluctuations of atomic energy levels due to the axion or scalar field dark matter are governed by a different statistical distribution as compared with ordinary noise which is often described by a normal Gaussian distribution, although other types of noise may also be present.

\subsection{Binning of experimental data}

Consider an experiment continuously measuring atomic or molecular energy level shifts with integration time $t_1$ and total duration of the experiment $t=Nt_1\gg t_1$. As a result, a series of experimental data are collected, $E_i$, $i=1,2,\ldots,N$, with mean $\bar E$ and standard deviation $\sigma$.

Assume that the mass $m$ of the axion (or scalar) field $\phi$ is known, and it satisfies the conditions (\ref{conditions}). As is argued in Sec.~\ref{SecI}, the amplitude $\phi_0$ of this field fluctuates with a typical coherence time $\tau \sim 10^6 T = 10^6 \pi/m$. This means that the fluctuations of energy shifts within one coherence time interval $\tau$ represent the noise in the detector, while the difference in mean energy shifts in different coherence time intervals is caused by stochastic fluctuations of the axion (or scalar) field amplitude. 

This suggests the following binning procedure of the experimental data set with the aim to suppress the noise. The total number of measurements $N$ is divided into $N_\text{bin}$ bins with $n_1$ data points in each bin such that $n_1 t_1 =\tau$. Thus, the energy shifts may be written as $E_{ik}$ with $i=1,2,\ldots,N_\text{bin}$ and $k=1,2,\ldots,n_1$. The energy shifts which fall within one bin may be averaged, $\bar E_i = \frac1{n_1}\sum_{k=1}^{n_1}E_{ik}$. As a result, the detector noise is averaged out, and fluctuations of $\bar E_i$ are mainly caused by axion or scalar field dark matter. As is shown in Sec.~\ref{SecI}, fluctuations of $\bar E_i$ should be governed (at least roughly) by the exponential distribution. This distribution is featured by the properties (\ref{sigmaE}) and (\ref{SK}). By checking that the energy level shifts $\bar E_i$ approximately satisfy these equations, one could conclude that the axion signal is present in the data.

The problem is, however, that the mass of the axion field and, hence, the coherence time are not known. Therefore, one has to look for the coherence time by diving the experimental data set into bins many times with different bin width $n_1$ from 1 to $N/2$. For each such binning one has to find the set of average energy shifts $\bar E_i$ and use it for calculation of standard deviation $\sigma$, skewness $S$ and kurtosis $K$. Thus, these statistical moments are functions of the bin width $n_1$. If the axion or scalar field signal is present in the experimental data, the values of these statistical moments should approximately satisfy the equations (\ref{sigmaE}) and (\ref{SK}) for certain $n_1$ such that $n_1 t_1 = \tau$. If such number $n_1$ is found, the axion mass is expressed as
\begin{equation}
    m \approx \kappa \frac{10^6 \pi}{n_1 t_1} \,.
    \label{m}
\end{equation}
It is usually assumed that the  coefficient $\kappa \sim 1$. Numerical simulations presented below give $\kappa\approx 0.55$. This coefficient depends on the specific dark matter velocity distribution $f(v)$.

This procedure may help finding an approximate value of the axion mass, which should serve as a motivation for further experiments to search for the axion particle with the mass near this value. It is reminiscent of the stacking procedure proposed in Ref.~\cite{Foster} for optimizing the storage and statistical analysis of experimental data.

\begin{figure}
    \centering
    \includegraphics[width=8.5cm]{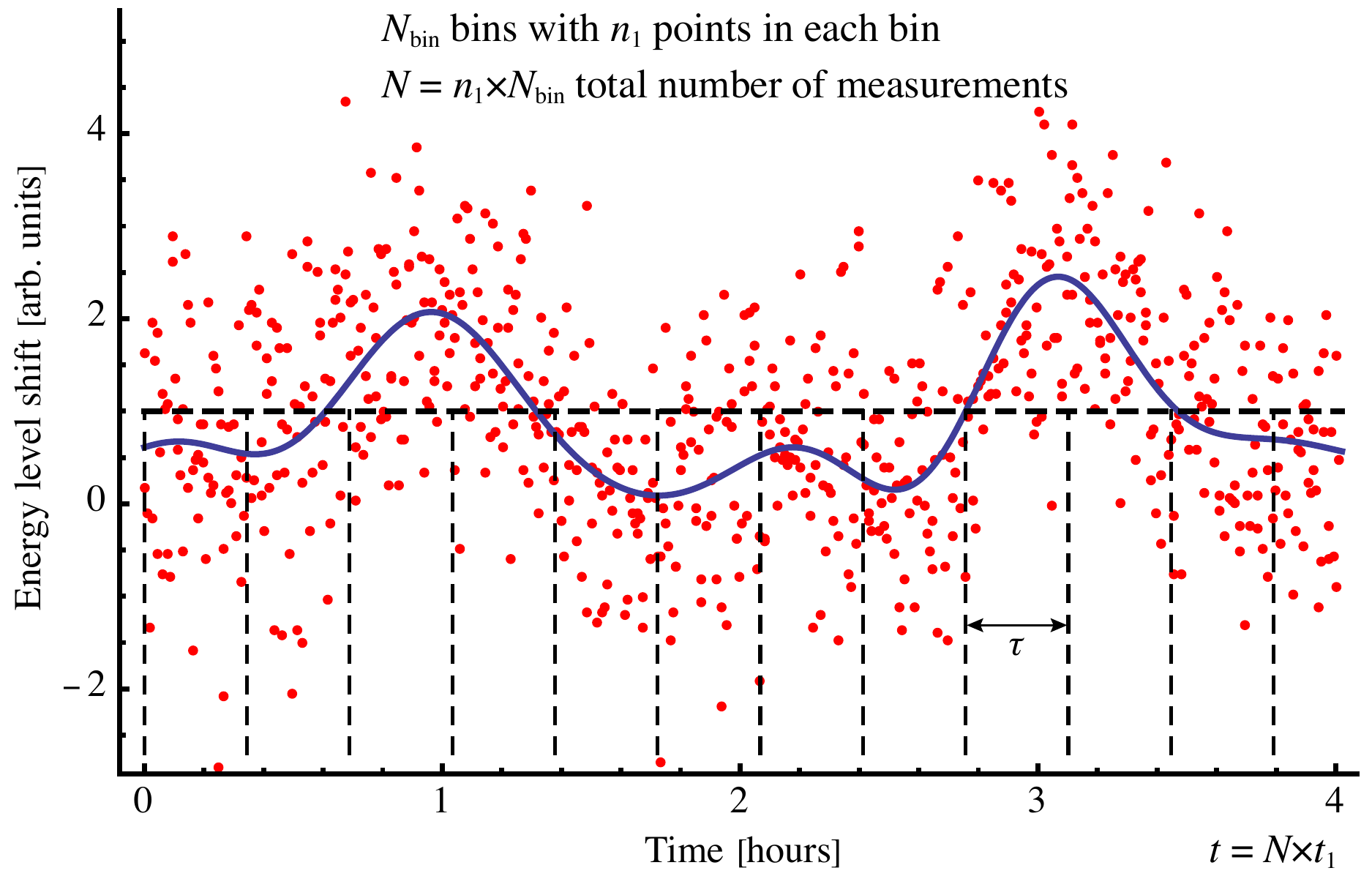}
    \caption{Illustration of binning of measurements of atomic energy level shift with atomic clocks to determine the axion coherence time. Blue curve represents contribution to the energy shift from stochastic fluctuations of the axion or scalar field amplitude. Red dots show random fluctuations of these energy level shift measurements due to white noise in the detector. Total measurement time $t=Nt_1$ is divided into time intervals equal to the duration of the scalar field coherence time $\tau\sim 10^6\pi/m$ such that each bin contains results of $n_1$ measurements. Energy shift measurements in each bin are distributed quasi Gaussian with mean $\bar E_i$ and dispersion $\sigma_i$. The energy shifts $\bar E_i$ in different bins follow the exponential distribution if the bin size corresponds to the dark matter coherence time.}
    \label{fig:scheme}
\end{figure}

\subsection{Numerical simulation}
\label{SecSimulations}

For an illustration of the binning procedure we perform a Monte-Carlo simulation of experimental data with $N=10^5$ energy level shifts generated as follows. First, we assume that the integration time is $t_1= 20.7$ s, and the total duration of the experiment is about 24 days, that is comparable with the parameters of the experiment \cite{HSi} considered is Sec.~\ref{SecHSi}. Then, we assume that the axion mass is $m=10^{-12}$ eV, so that the conditions (\ref{conditions}) are satisfied. With these parameters set, we generate a function $E_\phi(t) = C \langle \phi^2(t)\rangle$, where $\langle \phi^2(t)\rangle$ is given by Eq.~(\ref{26}) with $n=400$ pseudorandom phases $\varphi_j$ and Rayleigh distributed amplitudes $\alpha_j$. The coefficient $C$ is chosen, for simplicity, such that $\bar E_\phi \equiv \frac1t \int_0^t E_\phi(t)dt = 1$ in some units. Given this function, we check that it satisfies the condition (\ref{31}), as well as $S\approx 2$ and $K\approx 9$. Thus, it models the contribution to the atomic energy level shift due to ultralight axion dark matter.

Next, we generate a time series $E_i= E_\phi(t_i)+E_\text{noise}$, where $t_i = i\times t_1$, $i=1,2,\ldots,N$, and $E_\text{noise}$ represents a contribution to the energy level shift due to the noise in the detector. For simplicity, we consider a white Gaussian noise, although, more generally, other types of noise may be present in a real experiment. We assume that this noise is characterized by vanishing mean, $\overline{E_\text{noise}}=0$, and unit standard deviation $\sigma_\text{noise}=1$. This choice corresponds to commonly assumed regime with unit signal-to-noise ratio SNR=1. A part of generated data set (red dots) and the function $E_\phi(t)$ (blue curve) are shown in Fig.~\ref{fig:scheme}.

\begin{figure}
    \centering
    \includegraphics[width=8.5cm]{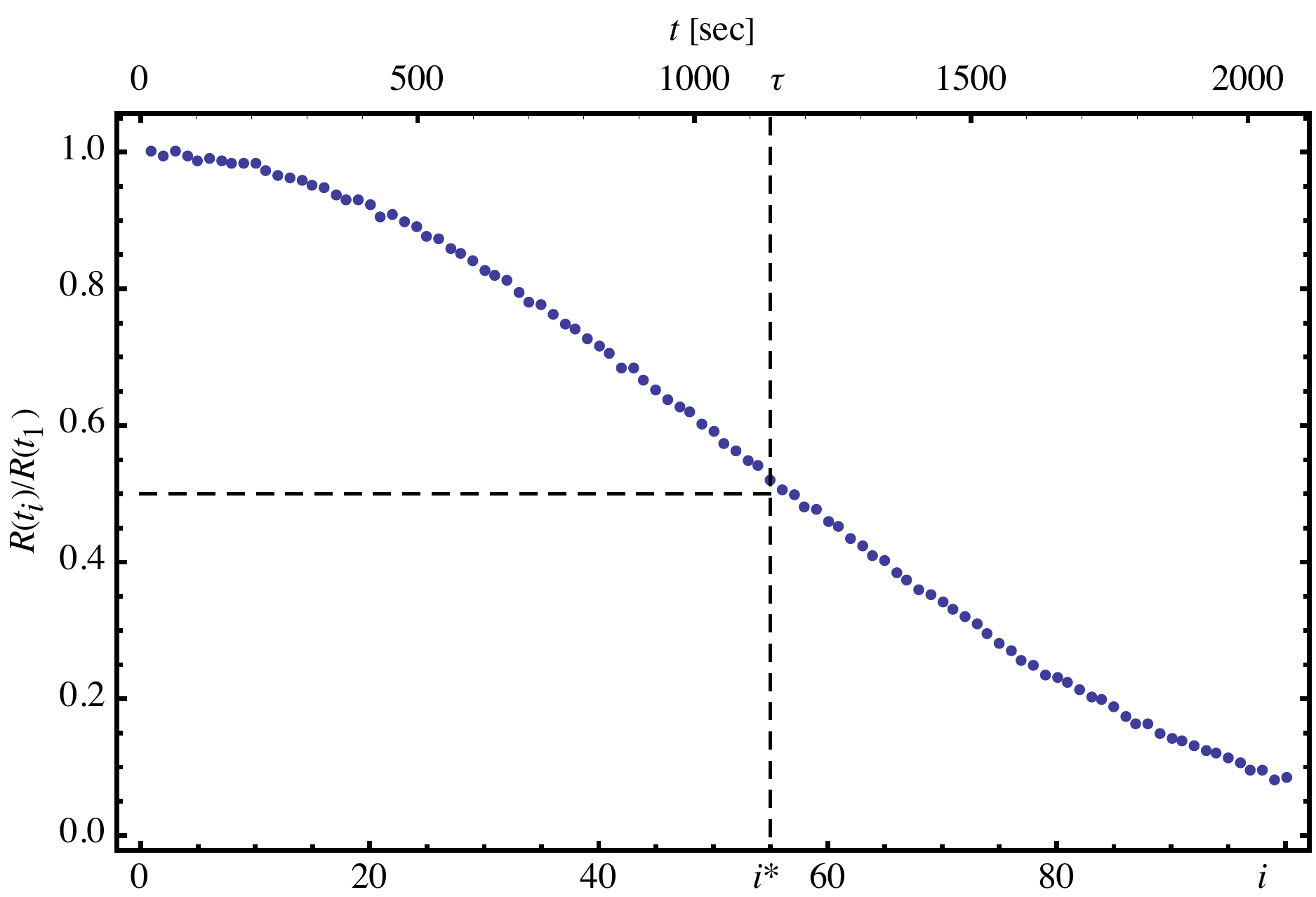}
    \caption{Relative values of the correlation function (\ref{Rsum}) for a mock data set considered in Sec.~\ref{SecSimulations}. This correlation function falls by half its initial value at $i^*=55$ that corresponds to the coherence time $\tau = i^* \times t_1 \approx 0.55 \cdot 10^6 \pi /m = 1150$ sec.}
    \label{fig:correlator}
\end{figure}

Assume now that the axion dark matter coherence time and the axion mass are not known, and they should be determined from the given mock data set $\{E_i\}$. First, one has to determine the coherence time with the use of Eq.~(\ref{Rtau}). For discrete time intervals $t_i= i\Delta t$, $1\leq i\ll N$, this correlation function may be represented as a sum:
\begin{equation}
    \label{Rsum}
    R(t_i) = \frac1{N-i} \sum_{j=1}^{N-i}(E_{i+j}-\bar E)(E_j-\bar E)\,.
\end{equation}
For the considered data set, the relative values of this correlation function are plotted in Fig.~\ref{fig:correlator}. This graph shows that the correlation function (\ref{Rsum}) falls by half its initial value at $i^*=55$ corresponding to $\tau = i^*\times t_1 \approx  0.55 \cdot 10^6 \pi /m =1150$ sec. 

\begin{figure}
    \centering
    \includegraphics[width=8.5cm]{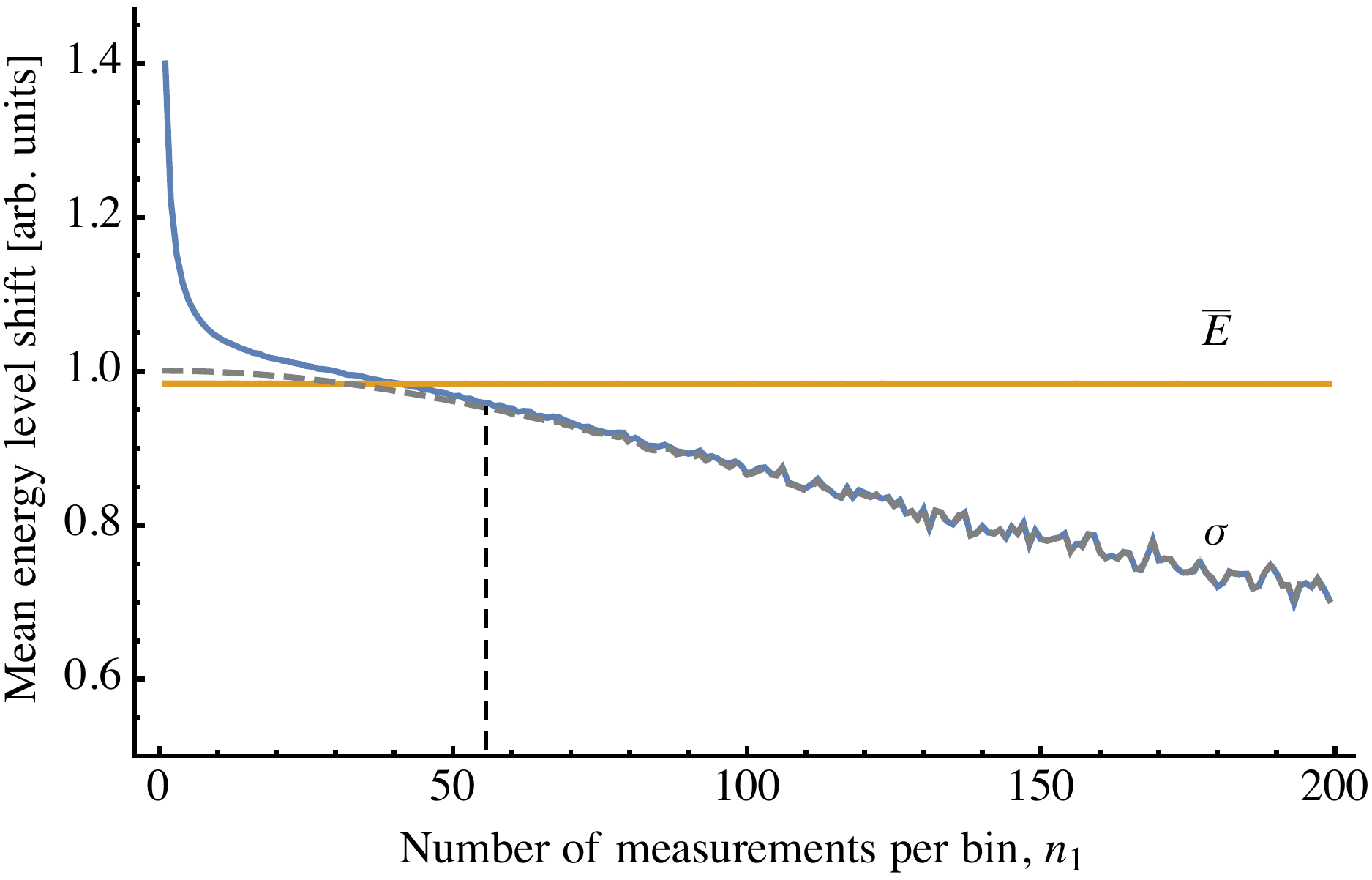}
    \caption{Simulation of the binning procedure for finding the axion coherence time. Horizontal axis measures the number of points per bin out of total $10^5$ simulated energy shift measurements. Vertical axis corresponds to mean atomic energy level shift $\bar E$ in arbitrary units. Blue curve represents the standard deviation $\sigma$ calculated for mean energy shifts $\bar E_i$ in each bin. This curve intersects the line $\sigma=\bar E$ near $n_1=55$ (dashed vertical line). This value may be used to estimate the axion amplitude coherence time $\tau = n_1 t_1$, with $t_1$ the integration time in one measurement. This value of the coherence time is in agreement with the definition (\ref{Rtau}). The gray dashed curve shows the values of standard deviation $\sigma$ with vanishing detector noise.}
    \label{fig:simulation}
\end{figure}

Next, we divide the data set $\{ E_i \}$ into $N_\text{bin}$ bins with $n_1$ points in each bin, as in Fig.~\ref{fig:scheme}. In each bin, the energy shifts are averaged, and values $\bar E_i$ are found, $i=1,2,\ldots,N_\text{bin}$. The mean $\bar E$ and standard deviation $\sigma$ are calculated using these $\bar E_i$. This calculation of $\bar E$ and $\sigma$ should be repeated for different values of the bin width $n_1$ in the interval $1 \leq n_1 \leq N/2$. For the mock data set under consideration, the values of $\bar E$ and $\sigma$ as functions of $n_1$ are plotted in Fig.~\ref{fig:simulation}. In this figure, the value of the standard deviation $\sigma$ decreases quasi monotonically with $n_1$ and crosses the line $\sigma=\bar E$ near $n_1 = i^*= 55$.  Given this value of $n_1$, the axion mass is estimated with Eq.~(\ref{m}). 

\begin{figure}
    \centering
    \includegraphics[width=8.5cm]{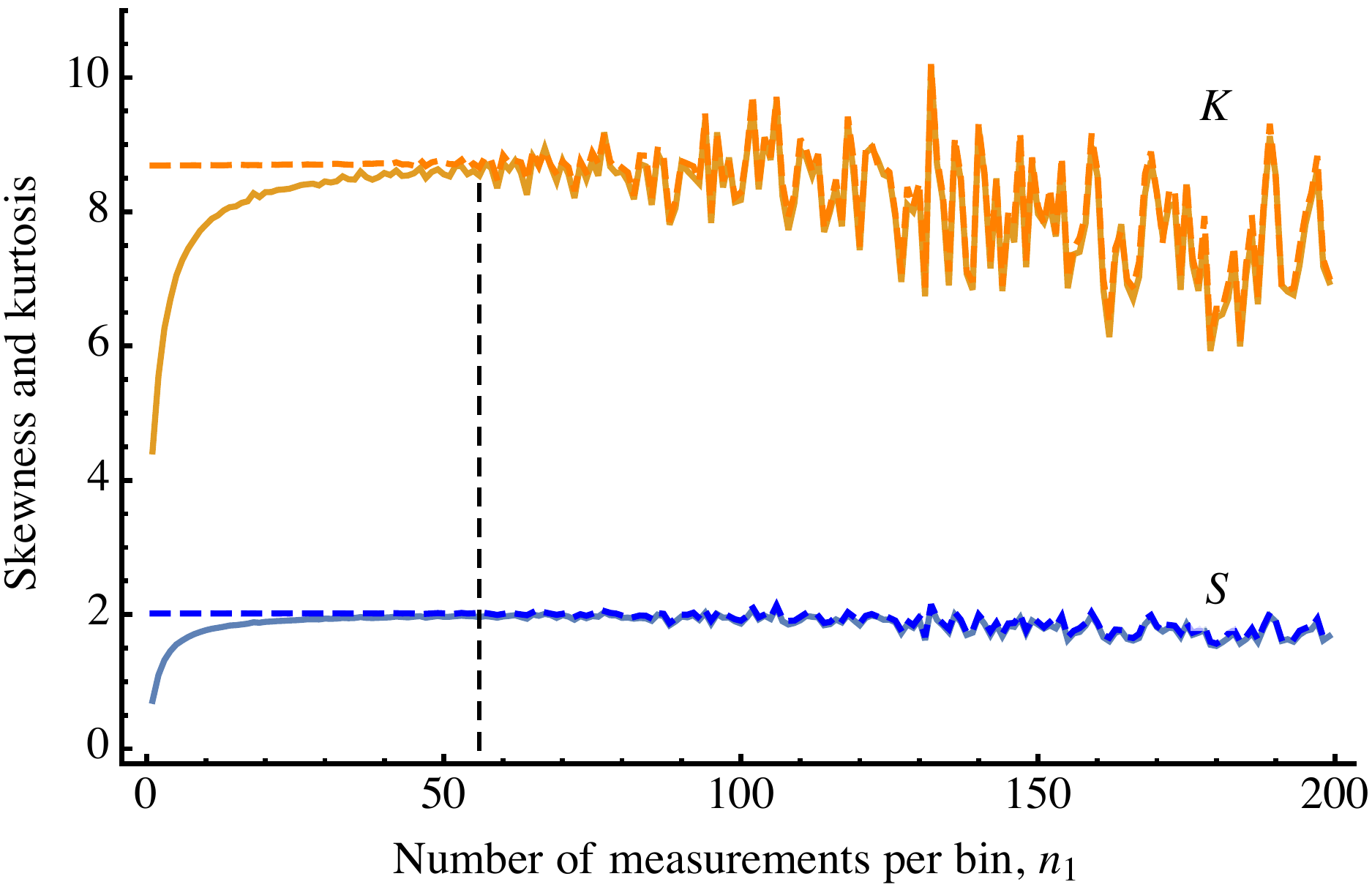}
    \caption{Same a in Fig.~\ref{fig:simulation}, but for the skewness $S$ and kurtosis $K$. Dashed horizontal lines show the values of the skewness and kurtosis with vanishing detector noise. At $n_1=55$ (vertical dashed line) skewness and kurtosis approximately reach their expected values (\ref{SK}) specific for exponential distributions. This helps determining the axion dark matter coherence time $\tau = n_1 t_1$.}
    \label{fig:simulationSK}
\end{figure}

Note that for small values of $n_1$, the standard deviation $\sigma$ in Fig.~\ref{fig:simulation} grows and reaches the value $\sigma\approx1.4$ at $n_1=1$. This behavior is natural because for small bin size the detector noise (modelled by Gaussian distribution with vanishing mean and unit standard deviation in the present case) dominates. The value $\sigma\approx 1.4$ is simply a standard deviation of the convolution of the two distributions (\ref{convolution}). Thus, our simulation confirms that for $n_1=1$ (no binning), the relation (\ref{sigmaLimit}) is satisfied, and experimentally measured standard deviation of the full data set may be used for extracting limits on the axion parameter space as in Sec.~\ref{SecLimits}.

Fig.~\ref{fig:simulation} demonstrates also that when the bin size is $n_1\gtrsim20$, the detector noise is averaged out, and the standard deviation appears very close to the gray dashed curve representing the values of the standard deviation with no detector noise (energy shift is fully produced by the axion signal). On the other hand, when $n_1>\tau/t_1$, the bin size becomes so large that not only detector noise, but also the axion field amplitude fluctuations average out, and the relation (\ref{31}) is strongly violated, $\sigma<\bar E$.

Additional information may be obtained from the plots of the skewness $S$ and kurtosis $K$ as functions of bin width $n_1$, see Fig.~\ref{fig:simulationSK}. These functions may be found in a similar way as the standard deviation considered above. Fig.~\ref{fig:simulationSK} demonstrates that when the bin width is small, the values of these parameters are much smaller than the predicted ones (\ref{SK}) because the contribution from the detector noise dominates in this regime. The values (\ref{SK}) are approximately reached when the bin size becomes $n_1 = \tau/t_1 =55$, that is indicated by vertical dashed line in Fig.~\ref{fig:simulationSK}. As a result, plots of skewness and kurtosis confirm that the energy level shifts $\bar E_i$ roughly follow the exponential distribution when the bin width corresponds to the axion coherence time interval. For real experiments, this could be a strong indication of presence of the axion signal in the experimental data.

To avoid misunderstanding, we must stress that we do not make ensemble averaging in our numerical model. The result is obtained with the fixed set of amplitudes $\alpha_j$ and phases $\Delta\varphi_{kj}$ in Eq. (\ref{26}). This is an adequate approach to simulate real experimental data. A different sets of amplitudes $\alpha_j$ and phases $\Delta\varphi_{kj}$ may produce  slightly different results but the conclusions will be the same.

The above simulation of the energy level shift qualitatively demonstrates the procedure for axion mass determination. This procedure may be applied when the noise in the detector is comparable with the expected energy level shift due to the axion dark matter; otherwise, if the noise is too high, the axion signal is totally washed out and the axion mass may not be found. However, if in a real experimental data a similar behaviour is observed, this would be a strong indication that variations of frequencies of atomic clocks are caused by fluctuations of the amplitude of the axion dark matter. It is very tempting to apply this procedure to the real data sets of experiments reported in Refs.~\cite{DyCs,YbCs,RbCs,HSi,Tretiak,Yb,DyQuartz,Huntemann}.


\section{Second-order pertubation theory correction to energy levels shift due to linear pseudoscalar interaction}
\label{Sec2Order}
Standard model spinor fields $\psi$, photon $F_{\mu\nu}$ and gluon $G^l_{\mu\nu}$ fields can have the following interaction vertices with a pseudoscalar field $\phi$:
\begin{equation}
\label{V1}
V= \frac{C_f}{f_a}  \partial_{\mu}  \phi  \bar \psi  \gamma_5 \gamma^{\mu}\psi + 
 C_{\gamma} \frac{\phi}{f_a} \tilde{F}^{\mu\nu} F_{\mu\nu} +  C_g \frac{\phi}{f_a} \tilde{G}^{l\,\mu\nu} G^l_{\mu\nu}\,.  
\end{equation}
Here $C_f$, $C_\gamma$ and $C_g$ are some dimensionless constants which are of order $O(1)$ for the QCD axion model, but are arbitrary for the general pseudoscalar (axion-like) particle. In particular, the last term in Eq.~(\ref{V1}) reduces to the QCD $\theta$-term (\ref{thetaQCD}) upon the substitution $C_g = g^2/(32\pi^2)$, or
\begin{equation}
\label{theta}
    \theta = \frac{32\pi^2 C_g\phi}{g^2 f_a}\,.
\end{equation}

In atoms and molecules, the interaction (\ref{V1}) cannot produce energy levels shifts in the first order of perturbation theory because the pseudoscalar field mixes the states of opposite parity if one neglects a small axion momentum corresponding to virialized dark matter particles in the standard dark matter halo model. Thus, non-trivial corrections to the energy levels $E_n$ start from the second order in the perturbation theory
\begin{equation}
\label{E2}
E^{(2)} = \sum_{n\ne0} \frac{\langle 0|V|n\rangle \langle n|V|0\rangle }{E_0- E_n}\,.
\end{equation}

The second-order energy corrections may be useful for studying variance of fluctuations of the energy shift averaged over the field oscillations. Such corrections may be significant in the cases of small energy denominators $E_0 -E_n$ in Eq.~(\ref{E2}) which is the case of close metastable states in Dy atom, polar molecules with rotational doublets, close levels in nuclear clock based on $^{229}$Th. In Dy atom and molecules, the energy interval $E_0- E_n$ may be reduced to zero by application of magnetic field. Near the level crossing the interval between the levels becomes linear in the perturbation, $|E_n-E_0|=2 |\langle 0|V|n\rangle |$. However, level widths and time-dependent perturbation $V$ make the problem more complicated.  We leave this problem for future study. 

The second-order energy correction (\ref{E2}) is quadratic in the pseudoscalar field amplitude, $E^{(2)}\propto \phi_0^2$.
Therefore, it is interesting to compare this second-order energy level shift with the effect of $\theta^2$ in pion mass (\ref{deltaPi}) discussed in Sec.~\ref{SecI}. Note that these two contributions to the energy level shift are produced by two independent mechanisms, although both originate from the same underlying axion-gluon interaction (\ref{thetaQCD}). Below we estimate the second-order energy level shift in $^{229}$Th nucleus caused by $CP$-violating pion-nucleon interaction $\sim \theta \pi \bar N N$ which was derived in Ref.~\cite{Witten} where neutron EDM due to QCD $\theta$-vacuum was calculated.

\subsection{Energy shift of $^{229}$Th isomeric state}

Consider, for example, the second-order contribution to the energy shift of the low-lying level $E \equiv \hbar\omega= 8.3$ eV of the nuclear clock transition in $^{229}$Th \cite{Seiferle2019}. This shift may be produced by $P,T$-violating nuclear forces with non-relativistic potential of the general form \cite{JETP1984}: 
\begin{equation}
\label{Vnuclear}
    V = \xi \vec\sigma \nabla V_s\,.
\end{equation} 
Here $\xi$ is a coupling constant, $\vec\sigma$ are the Pauli matrices corresponding to the spin of the nucleon and $V_s$ is the average nucleon-nucleus potential due to the strong interaction. 

In this paper, we consider a simple model where $V_s$ is given by an oscillator-type potential, 
\begin{equation}
\label{Vs}
V_s = V_0 (r^2/R^2-1)\,,
\end{equation}
with $V_0 \simeq 50$ MeV and the nuclear radius $R$. This potential vanishes on the boundary of the nucleus at $r=R$, and is negative inside the nucleus. Although this potential represents a crude nuclear model, it allows us to estimate analytically the second-order perturbative correction to the energy level shift of an isomeric state in $^{229}$Th. More accurate and sophisticated nuclear models would require numerical methods which are beyond the scope of this paper.

In Ref.~\cite{JETP1984} the constant $\xi$ was expressed via a dimensionless coupling $\eta$, $\xi=-2\times 10^{-21} \eta\,$cm, which was related with the QCD vacuum angle $\theta$ in Ref.~\cite{PospelovRitz,DeMille,deVries}: $\eta=4.4 \times 10^{5} \theta$. Indeed, the $P,T$-odd nuclear force (\ref{Vnuclear}) is dominated by the $\pi_0$ meson exchange \cite{DeMille}, while the coupling constants of $P,T$-odd pion-nucleon interaction were expressed via $\theta$ in the classic paper \cite{Witten}. Making use of Eq.~(\ref{theta}), we express $\xi$ via $C_g/f_a$:
\begin{equation}
    \xi = -8.8\times 10^{-16}\frac{32\pi^2 C_g\phi}{g^2 f_a}\text{cm}\,.
\end{equation}
Thus, the $CP$-odd potential (\ref{Vnuclear}) is first-order in the axion field $\phi = \theta f_a$, and the corresponding second-order energy correction (\ref{E2}) may be cast in the form 
\begin{equation} \label{E2psi}
E^{(2)} = \langle \delta \psi|V|\psi \rangle\,,
\end{equation}
where 
\begin{equation}
\label{deltapsi}
    |\delta\psi\rangle = \sum_{n\ne 0} \frac{ | n\rangle \langle n|V|\psi\rangle}{E_0 - E_n}
\end{equation}
is the first-order correction to the wave function $\psi$. In Ref.~\cite{JETP1984} this correction was found in the following simple form
\begin{equation}
 \delta \psi = \xi \vec\sigma \nabla \psi\,. 
\end{equation}
Substituting this function into Eq.~(\ref{E2psi}), integrating by parts and using commutation identities of Pauli matrices we find
\begin{equation}
    E^{(2)} = -3\frac{\xi^2 V_0}{R^2}-4\frac{\xi^2V_0}{R^2}\langle \vec l\cdot\vec s\rangle\,,
    \label{19}
\end{equation}
where $\vec l$ and $\vec s$ are the nuclear orbital momentum and spin operators, respectively.

Remember that the lowest transition frequency in $^{229}$Th is given by the difference between the energies of excited $3/2^+$ [633] and the ground $5/2^+$ [631] nuclear states, $\hbar\omega = E_{3/2^+}-E_{5/2^+}$. Eq.~(\ref{19}) allows us to find the frequency shift of this transition due to $P,T$-odd hadronic interaction (\ref{Vnuclear}), 
\begin{equation}
    \label{20}
\begin{aligned}
    \hbar\delta\omega &= -4\frac{V_0\xi^2}{R^2}\left(\langle \vec l\cdot\vec s\rangle_{3/2^+} -\langle \vec l\cdot\vec s\rangle_{5/2^+} \right) \\
    &=-8\frac{V_0\xi^2}{R^2}\,,
\end{aligned}
\end{equation}
where we made use of the identity $\langle \vec l\cdot\vec s\rangle_{3/2^+} -\langle \vec l\cdot\vec s\rangle_{5/2^+} =2$ \cite{Wiringa2}.

The nuclear charge radius of $^{229}$Th is $R\approx 7.43$ fm \cite{NuclearRadius}. Substituting this value into Eq.~(\ref{20}) we find the relative frequency shift
\begin{equation}
    \delta\omega/\omega \approx -68\theta^2 \,.
    \label{E2result}
\end{equation}
This result may be compared with the first-order energy shift due to the pion mass variation found in Ref.~\cite{KimPerez} (using  the calculation of the dependence of nuclear energy levels on pion mass from  Ref.~\cite{Wiringa2}): $ E^{(1)}/E= 2 \times 10 ^5 \delta m_{\pi}^2/ m_{\pi}^2\approx2\times 10^4 \theta^2$. Thus, the second-order energy shift (\ref{E2result}) is about 300 times smaller than the first-order contribution from pion mass $\theta$-dependence. This allows us to find the limits on the axion constant $f_a$ by re-scaling the corresponding limits from Ref.~\cite{KimPerez}. This limit is represented in Fig.~\ref{fig:limits} by the dashed line. We stress that the variation of frequency of nuclear clock considered in this section originates from the term $\cos(2\omega t)$ in $\phi^2$ rather than from fluctuations of the (pseudo)scalar field amplitude.


\section{Summary}
\label{SecSummary}
In this paper, we found two new effects in the QCD axion model which contribute to variations of fundamental constants. 

The first effect appears from quadratic axion-nucleon interaction (\ref{1}) originating from the quadratic dependence of the pion mass on the axion (\ref{deltaPi}). This effect may be observed via variations of frequencies of atomic clocks due to fluctuations of the (pseudo)scalar field amplitude. We show that these variations of frequencies may be identified with variance of  measured fluctuations of transition frequencies in the atomic clocks, see Eq.~(\ref{sigmaE}). This allows one to explore the region of QCD axion masses satisfying Eq.~(\ref{conditions}). By re-purposing correspondingly the results of the experiments \cite{RbCs} and \cite{HSi} we found new laboratory limits on the axion decay constant $f_a$ in the range $2.4\times 10^{-17}\text{ eV}\lesssim m \lesssim 10^{-13}\text{ eV}$. 

We propose also a procedure which, in principle, could help finding signatures of the axion signal in the energy level shift measurements in atomic clock experiments. This procedure includes binning of the data, averaging of the data points inside each bin and calculations of standard deviation $\sigma$, skewness $S$, and kurtosis $K$ for the resulting distribution. The averaging procedure allows one to suppress the detector noise while keeping the axion signal if it is present. If the axion signal has magnitude comparable or bigger than the noise, graphs of the correlation function $R(t)$ and $\sigma$, $S$ and $K$ as a function of the bin width allows one to approximately find the value of the coherence time $\tau$, axion mass $m c^2 \sim 1.7 \times  10^6 \hbar/ \tau $ and average energy shift $E$ produced by the axion field, see Figs.~\ref{fig:correlator}, \ref{fig:simulation} and \ref{fig:simulationSK}.

The other effect originates from the second-order perturbative correction to the energy level shift due to linear-in-$\phi$ interaction. Since this effect is expected to be small, it may manifest itself only in extremely accurate frequency measurements with future technology based on nuclear clocks. We estimated this shift for the low-energy nuclear transition in $^{229}$Th and found the projected limits from the expected sensitivity of such nuclear clocks, see Fig.~\ref{fig:limits}.

\vspace{2mm}
\textit{Acknowledgements.}--- 
We are indebted to the Referee for pointing out that the Rayleigh distribution for the axion field amplitude results in non-trivial values of higher statistical moments (\ref{SK}) and for proposing the idea of binning procedure for searches of possible signatures of the axion dark matter in atomic clock experiments.
We are grateful also to Dmitry Budker and Yevgeny Stadnik for informing us about Ref.~\cite{GNOMEdE} and to Melina Filzinger and Nils Huntemann for valuable comments. The work was supported by the Australian Research Council Grants No.\ DP230101058 and DP200100150.

%


\begin{thebibliography}{53}%
\makeatletter
\providecommand \@ifxundefined [1]{%
 \@ifx{#1\undefined}
}%
\providecommand \@ifnum [1]{%
 \ifnum #1\expandafter \@firstoftwo
 \else \expandafter \@secondoftwo
 \fi
}%
\providecommand \@ifx [1]{%
 \ifx #1\expandafter \@firstoftwo
 \else \expandafter \@secondoftwo
 \fi
}%
\providecommand \natexlab [1]{#1}%
\providecommand \enquote  [1]{``#1''}%
\providecommand \bibnamefont  [1]{#1}%
\providecommand \bibfnamefont [1]{#1}%
\providecommand \citenamefont [1]{#1}%
\providecommand \href@noop [0]{\@secondoftwo}%
\providecommand \href [0]{\begingroup \@sanitize@url \@href}%
\providecommand \@href[1]{\@@startlink{#1}\@@href}%
\providecommand \@@href[1]{\endgroup#1\@@endlink}%
\providecommand \@sanitize@url [0]{\catcode `\\12\catcode `\$12\catcode
  `\&12\catcode `\#12\catcode `\^12\catcode `\_12\catcode `\%12\relax}%
\providecommand \@@startlink[1]{}%
\providecommand \@@endlink[0]{}%
\providecommand \url  [0]{\begingroup\@sanitize@url \@url }%
\providecommand \@url [1]{\endgroup\@href {#1}{\urlprefix }}%
\providecommand \urlprefix  [0]{URL }%
\providecommand \Eprint [0]{\href }%
\providecommand \doibase [0]{https://doi.org/}%
\providecommand \selectlanguage [0]{\@gobble}%
\providecommand \bibinfo  [0]{\@secondoftwo}%
\providecommand \bibfield  [0]{\@secondoftwo}%
\providecommand \translation [1]{[#1]}%
\providecommand \BibitemOpen [0]{}%
\providecommand \bibitemStop [0]{}%
\providecommand \bibitemNoStop [0]{.\EOS\space}%
\providecommand \EOS [0]{\spacefactor3000\relax}%
\providecommand \BibitemShut  [1]{\csname bibitem#1\endcsname}%
\let\auto@bib@innerbib\@empty
\bibitem [{\citenamefont {Preskill}\ \emph {et~al.}(1983)\citenamefont
  {Preskill}, \citenamefont {Wise},\ and\ \citenamefont {Wilczek}}]{Preskill}%
  \BibitemOpen
  \bibfield  {author} {\bibinfo {author} {\bibfnamefont {J.}~\bibnamefont
  {Preskill}}, \bibinfo {author} {\bibfnamefont {M.~B.}\ \bibnamefont {Wise}},\
  and\ \bibinfo {author} {\bibfnamefont {F.}~\bibnamefont {Wilczek}},\ }\href
  {https://doi.org/https://doi.org/10.1016/0370-2693(83)90637-8} {\bibfield
  {journal} {\bibinfo  {journal} {Phys. Lett. B}\ }\textbf {\bibinfo {volume}
  {120}},\ \bibinfo {pages} {127} (\bibinfo {year} {1983})}\BibitemShut
  {NoStop}%
\bibitem [{\citenamefont {Abbott}\ and\ \citenamefont
  {Sikivie}(1983)}]{Abbott}%
  \BibitemOpen
  \bibfield  {author} {\bibinfo {author} {\bibfnamefont {L.}~\bibnamefont
  {Abbott}}\ and\ \bibinfo {author} {\bibfnamefont {P.}~\bibnamefont
  {Sikivie}},\ }\href
  {https://doi.org/https://doi.org/10.1016/0370-2693(83)90638-X} {\bibfield
  {journal} {\bibinfo  {journal} {Phys. Lett. B}\ }\textbf {\bibinfo {volume}
  {120}},\ \bibinfo {pages} {133} (\bibinfo {year} {1983})}\BibitemShut
  {NoStop}%
\bibitem [{\citenamefont {Dine}\ and\ \citenamefont {Fischler}(1983)}]{Dine}%
  \BibitemOpen
  \bibfield  {author} {\bibinfo {author} {\bibfnamefont {M.}~\bibnamefont
  {Dine}}\ and\ \bibinfo {author} {\bibfnamefont {W.}~\bibnamefont
  {Fischler}},\ }\href
  {https://doi.org/https://doi.org/10.1016/0370-2693(83)90639-1} {\bibfield
  {journal} {\bibinfo  {journal} {Phys. Lett. B}\ }\textbf {\bibinfo {volume}
  {120}},\ \bibinfo {pages} {137} (\bibinfo {year} {1983})}\BibitemShut
  {NoStop}%
\bibitem [{\citenamefont {Drukier}\ \emph {et~al.}(1986)\citenamefont
  {Drukier}, \citenamefont {Freese},\ and\ \citenamefont {Spergel}}]{DMhalo}%
  \BibitemOpen
  \bibfield  {author} {\bibinfo {author} {\bibfnamefont {A.~K.}\ \bibnamefont
  {Drukier}}, \bibinfo {author} {\bibfnamefont {K.}~\bibnamefont {Freese}},\
  and\ \bibinfo {author} {\bibfnamefont {D.~N.}\ \bibnamefont {Spergel}},\
  }\href {https://doi.org/10.1103/PhysRevD.33.3495} {\bibfield  {journal}
  {\bibinfo  {journal} {Phys. Rev. D}\ }\textbf {\bibinfo {volume} {33}},\
  \bibinfo {pages} {3495} (\bibinfo {year} {1986})}\BibitemShut {NoStop}%
\bibitem [{\citenamefont {Pillepich}\ \emph {et~al.}(2014)\citenamefont
  {Pillepich}, \citenamefont {Kuhlen}, \citenamefont {Guedes},\ and\
  \citenamefont {Madau}}]{halo1}%
  \BibitemOpen
  \bibfield  {author} {\bibinfo {author} {\bibfnamefont {A.}~\bibnamefont
  {Pillepich}}, \bibinfo {author} {\bibfnamefont {M.}~\bibnamefont {Kuhlen}},
  \bibinfo {author} {\bibfnamefont {J.}~\bibnamefont {Guedes}},\ and\ \bibinfo
  {author} {\bibfnamefont {P.}~\bibnamefont {Madau}},\ }\href
  {https://doi.org/10.1088/0004-637X/784/2/161} {\bibfield  {journal} {\bibinfo
   {journal} {Astrophys. J.}\ }\textbf {\bibinfo {volume} {784}},\ \bibinfo
  {pages} {161} (\bibinfo {year} {2014})}\BibitemShut {NoStop}%
\bibitem [{\citenamefont {Evans}\ \emph {et~al.}(2019)\citenamefont {Evans},
  \citenamefont {O'Hare},\ and\ \citenamefont {McCabe}}]{DMhalo2}%
  \BibitemOpen
  \bibfield  {author} {\bibinfo {author} {\bibfnamefont {N.~W.}\ \bibnamefont
  {Evans}}, \bibinfo {author} {\bibfnamefont {C.~A.~J.}\ \bibnamefont
  {O'Hare}},\ and\ \bibinfo {author} {\bibfnamefont {C.}~\bibnamefont
  {McCabe}},\ }\href {https://doi.org/10.1103/PhysRevD.99.023012} {\bibfield
  {journal} {\bibinfo  {journal} {Phys. Rev. D}\ }\textbf {\bibinfo {volume}
  {99}},\ \bibinfo {pages} {023012} (\bibinfo {year} {2019})}\BibitemShut
  {NoStop}%
\bibitem [{\citenamefont {Arvanitaki}\ \emph {et~al.}(2015)\citenamefont
  {Arvanitaki}, \citenamefont {Huang},\ and\ \citenamefont
  {Van~Tilburg}}]{Arvanitaki}%
  \BibitemOpen
  \bibfield  {author} {\bibinfo {author} {\bibfnamefont {A.}~\bibnamefont
  {Arvanitaki}}, \bibinfo {author} {\bibfnamefont {J.}~\bibnamefont {Huang}},\
  and\ \bibinfo {author} {\bibfnamefont {K.}~\bibnamefont {Van~Tilburg}},\
  }\href {https://doi.org/10.1103/PhysRevD.91.015015} {\bibfield  {journal}
  {\bibinfo  {journal} {Phys. Rev. D}\ }\textbf {\bibinfo {volume} {91}},\
  \bibinfo {pages} {015015} (\bibinfo {year} {2015})}\BibitemShut {NoStop}%
\bibitem [{\citenamefont {Stadnik}\ and\ \citenamefont
  {Flambaum}(2015)}]{Stadnik}%
  \BibitemOpen
  \bibfield  {author} {\bibinfo {author} {\bibfnamefont {Y.~V.}\ \bibnamefont
  {Stadnik}}\ and\ \bibinfo {author} {\bibfnamefont {V.~V.}\ \bibnamefont
  {Flambaum}},\ }\href {https://doi.org/10.1103/PhysRevLett.115.201301}
  {\bibfield  {journal} {\bibinfo  {journal} {Phys. Rev. Lett.}\ }\textbf
  {\bibinfo {volume} {115}},\ \bibinfo {pages} {201301} (\bibinfo {year}
  {2015})}\BibitemShut {NoStop}%
\bibitem [{\citenamefont {Stadnik}\ and\ \citenamefont
  {Flambaum}(2016)}]{Stadnik2}%
  \BibitemOpen
  \bibfield  {author} {\bibinfo {author} {\bibfnamefont {Y.~V.}\ \bibnamefont
  {Stadnik}}\ and\ \bibinfo {author} {\bibfnamefont {V.~V.}\ \bibnamefont
  {Flambaum}},\ }\href {https://doi.org/10.1103/PhysRevA.94.022111} {\bibfield
  {journal} {\bibinfo  {journal} {Phys. Rev. A}\ }\textbf {\bibinfo {volume}
  {94}},\ \bibinfo {pages} {022111} (\bibinfo {year} {2016})}\BibitemShut
  {NoStop}%
\bibitem [{\citenamefont {Van~Tilburg}\ \emph {et~al.}(2015)\citenamefont
  {Van~Tilburg}, \citenamefont {Leefer}, \citenamefont {Bougas},\ and\
  \citenamefont {Budker}}]{DyCs}%
  \BibitemOpen
  \bibfield  {author} {\bibinfo {author} {\bibfnamefont {K.}~\bibnamefont
  {Van~Tilburg}}, \bibinfo {author} {\bibfnamefont {N.}~\bibnamefont {Leefer}},
  \bibinfo {author} {\bibfnamefont {L.}~\bibnamefont {Bougas}},\ and\ \bibinfo
  {author} {\bibfnamefont {D.}~\bibnamefont {Budker}},\ }\href
  {https://doi.org/10.1103/PhysRevLett.115.011802} {\bibfield  {journal}
  {\bibinfo  {journal} {Phys. Rev. Lett.}\ }\textbf {\bibinfo {volume} {115}},\
  \bibinfo {pages} {011802} (\bibinfo {year} {2015})}\BibitemShut {NoStop}%
\bibitem [{\citenamefont {Hees}\ \emph {et~al.}(2016)\citenamefont {Hees},
  \citenamefont {Gu\'ena}, \citenamefont {Abgrall}, \citenamefont {Bize},\ and\
  \citenamefont {Wolf}}]{RbCs}%
  \BibitemOpen
  \bibfield  {author} {\bibinfo {author} {\bibfnamefont {A.}~\bibnamefont
  {Hees}}, \bibinfo {author} {\bibfnamefont {J.}~\bibnamefont {Gu\'ena}},
  \bibinfo {author} {\bibfnamefont {M.}~\bibnamefont {Abgrall}}, \bibinfo
  {author} {\bibfnamefont {S.}~\bibnamefont {Bize}},\ and\ \bibinfo {author}
  {\bibfnamefont {P.}~\bibnamefont {Wolf}},\ }\href
  {https://doi.org/10.1103/PhysRevLett.117.061301} {\bibfield  {journal}
  {\bibinfo  {journal} {Phys. Rev. Lett.}\ }\textbf {\bibinfo {volume} {117}},\
  \bibinfo {pages} {061301} (\bibinfo {year} {2016})}\BibitemShut {NoStop}%
\bibitem [{\citenamefont {Kobayashi}\ \emph {et~al.}(2022)\citenamefont
  {Kobayashi}, \citenamefont {Takamizawa}, \citenamefont {Akamatsu},
  \citenamefont {Kawasaki}, \citenamefont {Nishiyama}, \citenamefont {Hosaka},
  \citenamefont {Hisai}, \citenamefont {Wada}, \citenamefont {Inaba},
  \citenamefont {Tanabe},\ and\ \citenamefont {Yasuda}}]{YbCs}%
  \BibitemOpen
  \bibfield  {author} {\bibinfo {author} {\bibfnamefont {T.}~\bibnamefont
  {Kobayashi}}, \bibinfo {author} {\bibfnamefont {A.}~\bibnamefont
  {Takamizawa}}, \bibinfo {author} {\bibfnamefont {D.}~\bibnamefont
  {Akamatsu}}, \bibinfo {author} {\bibfnamefont {A.}~\bibnamefont {Kawasaki}},
  \bibinfo {author} {\bibfnamefont {A.}~\bibnamefont {Nishiyama}}, \bibinfo
  {author} {\bibfnamefont {K.}~\bibnamefont {Hosaka}}, \bibinfo {author}
  {\bibfnamefont {Y.}~\bibnamefont {Hisai}}, \bibinfo {author} {\bibfnamefont
  {M.}~\bibnamefont {Wada}}, \bibinfo {author} {\bibfnamefont {H.}~\bibnamefont
  {Inaba}}, \bibinfo {author} {\bibfnamefont {T.}~\bibnamefont {Tanabe}},\ and\
  \bibinfo {author} {\bibfnamefont {M.}~\bibnamefont {Yasuda}},\ }\href
  {https://doi.org/10.1103/PhysRevLett.129.241301} {\bibfield  {journal}
  {\bibinfo  {journal} {Phys. Rev. Lett.}\ }\textbf {\bibinfo {volume} {129}},\
  \bibinfo {pages} {241301} (\bibinfo {year} {2022})}\BibitemShut {NoStop}%
\bibitem [{\citenamefont {Kennedy}\ \emph {et~al.}(2020)\citenamefont {Kennedy}
  \emph {et~al.}}]{HSi}%
  \BibitemOpen
  \bibfield  {author} {\bibinfo {author} {\bibfnamefont {C.~J.}\ \bibnamefont
  {Kennedy}} \emph {et~al.},\ }\href
  {https://doi.org/10.1103/PhysRevLett.125.201302} {\bibfield  {journal}
  {\bibinfo  {journal} {Phys. Rev. Lett.}\ }\textbf {\bibinfo {volume} {125}},\
  \bibinfo {pages} {201302} (\bibinfo {year} {2020})}\BibitemShut {NoStop}%
\bibitem [{\citenamefont {Tretiak}\ \emph {et~al.}(2022)\citenamefont {Tretiak}
  \emph {et~al.}}]{Tretiak}%
  \BibitemOpen
  \bibfield  {author} {\bibinfo {author} {\bibfnamefont {O.}~\bibnamefont
  {Tretiak}} \emph {et~al.},\ }\href
  {https://doi.org/10.1103/PhysRevLett.129.031301} {\bibfield  {journal}
  {\bibinfo  {journal} {Phys. Rev. Lett.}\ }\textbf {\bibinfo {volume} {129}},\
  \bibinfo {pages} {031301} (\bibinfo {year} {2022})}\BibitemShut {NoStop}%
\bibitem [{\citenamefont {Banerjee}\ \emph {et~al.}(2023)\citenamefont
  {Banerjee}, \citenamefont {Budker}, \citenamefont {Filzinger}, \citenamefont
  {Huntemann}, \citenamefont {Paz}, \citenamefont {Perez}, \citenamefont
  {Porsev},\ and\ \citenamefont {Safronova}}]{Yb}%
  \BibitemOpen
  \bibfield  {author} {\bibinfo {author} {\bibfnamefont {A.}~\bibnamefont
  {Banerjee}}, \bibinfo {author} {\bibfnamefont {D.}~\bibnamefont {Budker}},
  \bibinfo {author} {\bibfnamefont {M.}~\bibnamefont {Filzinger}}, \bibinfo
  {author} {\bibfnamefont {N.}~\bibnamefont {Huntemann}}, \bibinfo {author}
  {\bibfnamefont {G.}~\bibnamefont {Paz}}, \bibinfo {author} {\bibfnamefont
  {G.}~\bibnamefont {Perez}}, \bibinfo {author} {\bibfnamefont
  {S.}~\bibnamefont {Porsev}},\ and\ \bibinfo {author} {\bibfnamefont
  {M.}~\bibnamefont {Safronova}},\ }\href@noop {} {\bibinfo {title}
  {{Oscillating nuclear charge radii as sensors for ultralight dark matter}}}
  (\bibinfo {year} {2023}),\ \Eprint {https://arxiv.org/abs/2301.10784}
  {arXiv:2301.10784 [hep-ph]} \BibitemShut {NoStop}%
\bibitem [{\citenamefont {Filzinger}\ \emph {et~al.}(2023)\citenamefont
  {Filzinger}, \citenamefont {D\"orscher}, \citenamefont {Lange}, \citenamefont
  {Klose}, \citenamefont {Steinel}, \citenamefont {Benkler}, \citenamefont
  {Peik}, \citenamefont {Lisdat},\ and\ \citenamefont {Huntemann}}]{Huntemann}%
  \BibitemOpen
  \bibfield  {author} {\bibinfo {author} {\bibfnamefont {M.}~\bibnamefont
  {Filzinger}}, \bibinfo {author} {\bibfnamefont {S.}~\bibnamefont
  {D\"orscher}}, \bibinfo {author} {\bibfnamefont {R.}~\bibnamefont {Lange}},
  \bibinfo {author} {\bibfnamefont {J.}~\bibnamefont {Klose}}, \bibinfo
  {author} {\bibfnamefont {M.}~\bibnamefont {Steinel}}, \bibinfo {author}
  {\bibfnamefont {E.}~\bibnamefont {Benkler}}, \bibinfo {author} {\bibfnamefont
  {E.}~\bibnamefont {Peik}}, \bibinfo {author} {\bibfnamefont {C.}~\bibnamefont
  {Lisdat}},\ and\ \bibinfo {author} {\bibfnamefont {N.}~\bibnamefont
  {Huntemann}},\ }\href {https://doi.org/10.1103/PhysRevLett.130.253001}
  {\bibfield  {journal} {\bibinfo  {journal} {Phys. Rev. Lett.}\ }\textbf
  {\bibinfo {volume} {130}},\ \bibinfo {pages} {253001} (\bibinfo {year}
  {2023})}\BibitemShut {NoStop}%
\bibitem [{\citenamefont {Zhang}\ \emph {et~al.}(2023)\citenamefont {Zhang},
  \citenamefont {Banerjee}, \citenamefont {Leyser}, \citenamefont {Perez},
  \citenamefont {Schiller}, \citenamefont {Budker},\ and\ \citenamefont
  {Antypas}}]{DyQuartz}%
  \BibitemOpen
  \bibfield  {author} {\bibinfo {author} {\bibfnamefont {X.}~\bibnamefont
  {Zhang}}, \bibinfo {author} {\bibfnamefont {A.}~\bibnamefont {Banerjee}},
  \bibinfo {author} {\bibfnamefont {M.}~\bibnamefont {Leyser}}, \bibinfo
  {author} {\bibfnamefont {G.}~\bibnamefont {Perez}}, \bibinfo {author}
  {\bibfnamefont {S.}~\bibnamefont {Schiller}}, \bibinfo {author}
  {\bibfnamefont {D.}~\bibnamefont {Budker}},\ and\ \bibinfo {author}
  {\bibfnamefont {D.}~\bibnamefont {Antypas}},\ }\href
  {https://doi.org/10.1103/PhysRevLett.130.251002} {\bibfield  {journal}
  {\bibinfo  {journal} {Phys. Rev. Lett.}\ }\textbf {\bibinfo {volume} {130}},\
  \bibinfo {pages} {251002} (\bibinfo {year} {2023})}\BibitemShut {NoStop}%
\bibitem [{\citenamefont {Sherrill}\ \emph {et~al.}(2023)\citenamefont
  {Sherrill} \emph {et~al.}}]{NPL}%
  \BibitemOpen
  \bibfield  {author} {\bibinfo {author} {\bibfnamefont {N.}~\bibnamefont
  {Sherrill}} \emph {et~al.},\ }\href@noop {} {\bibinfo {title} {{Analysis of
  atomic-clock data to constrain variations of fundamental constants}}}
  (\bibinfo {year} {2023}),\ \Eprint {https://arxiv.org/abs/2302.04565}
  {arXiv:2302.04565 [physics.atom-ph]} \BibitemShut {NoStop}%
\bibitem [{\citenamefont {Dzuba}\ \emph
  {et~al.}(1999{\natexlab{a}})\citenamefont {Dzuba}, \citenamefont {Flambaum},\
  and\ \citenamefont {Webb}}]{PRLWebb}%
  \BibitemOpen
  \bibfield  {author} {\bibinfo {author} {\bibfnamefont {V.~A.}\ \bibnamefont
  {Dzuba}}, \bibinfo {author} {\bibfnamefont {V.~V.}\ \bibnamefont
  {Flambaum}},\ and\ \bibinfo {author} {\bibfnamefont {J.~K.}\ \bibnamefont
  {Webb}},\ }\href {https://doi.org/10.1103/PhysRevLett.82.888} {\bibfield
  {journal} {\bibinfo  {journal} {Phys. Rev. Lett.}\ }\textbf {\bibinfo
  {volume} {82}},\ \bibinfo {pages} {888} (\bibinfo {year}
  {1999}{\natexlab{a}})}\BibitemShut {NoStop}%
\bibitem [{\citenamefont {Dzuba}\ \emph
  {et~al.}(1999{\natexlab{b}})\citenamefont {Dzuba}, \citenamefont {Flambaum},\
  and\ \citenamefont {Webb}}]{PRAWebb}%
  \BibitemOpen
  \bibfield  {author} {\bibinfo {author} {\bibfnamefont {V.~A.}\ \bibnamefont
  {Dzuba}}, \bibinfo {author} {\bibfnamefont {V.~V.}\ \bibnamefont
  {Flambaum}},\ and\ \bibinfo {author} {\bibfnamefont {J.~K.}\ \bibnamefont
  {Webb}},\ }\href {https://doi.org/10.1103/PhysRevA.59.230} {\bibfield
  {journal} {\bibinfo  {journal} {Phys. Rev. A}\ }\textbf {\bibinfo {volume}
  {59}},\ \bibinfo {pages} {230} (\bibinfo {year}
  {1999}{\natexlab{b}})}\BibitemShut {NoStop}%
\bibitem [{\citenamefont {Flambaum}\ and\ \citenamefont
  {Dzuba}(2009)}]{CanJPh}%
  \BibitemOpen
  \bibfield  {author} {\bibinfo {author} {\bibfnamefont {V.~V.}\ \bibnamefont
  {Flambaum}}\ and\ \bibinfo {author} {\bibfnamefont {V.~A.}\ \bibnamefont
  {Dzuba}},\ }\href {https://doi.org/10.1139/p08-072} {\bibfield  {journal}
  {\bibinfo  {journal} {Can. J. Phys.}\ }\textbf {\bibinfo {volume} {87}},\
  \bibinfo {pages} {25} (\bibinfo {year} {2009})}\BibitemShut {NoStop}%
\bibitem [{\citenamefont {Flambaum}\ and\ \citenamefont
  {Tedesco}(2006)}]{Tedesco}%
  \BibitemOpen
  \bibfield  {author} {\bibinfo {author} {\bibfnamefont {V.~V.}\ \bibnamefont
  {Flambaum}}\ and\ \bibinfo {author} {\bibfnamefont {A.~F.}\ \bibnamefont
  {Tedesco}},\ }\href {https://doi.org/10.1103/PhysRevC.73.055501} {\bibfield
  {journal} {\bibinfo  {journal} {Phys. Rev. C}\ }\textbf {\bibinfo {volume}
  {73}},\ \bibinfo {pages} {055501} (\bibinfo {year} {2006})}\BibitemShut
  {NoStop}%
\bibitem [{\citenamefont {Pa\v{s}teka}\ \emph {et~al.}(2019)\citenamefont
  {Pa\v{s}teka}, \citenamefont {Hao}, \citenamefont {Borschevsky},
  \citenamefont {Flambaum},\ and\ \citenamefont {Schwerdtfeger}}]{Borschevsky}%
  \BibitemOpen
  \bibfield  {author} {\bibinfo {author} {\bibfnamefont {L.~F.}\ \bibnamefont
  {Pa\v{s}teka}}, \bibinfo {author} {\bibfnamefont {Y.}~\bibnamefont {Hao}},
  \bibinfo {author} {\bibfnamefont {A.}~\bibnamefont {Borschevsky}}, \bibinfo
  {author} {\bibfnamefont {V.~V.}\ \bibnamefont {Flambaum}},\ and\ \bibinfo
  {author} {\bibfnamefont {P.}~\bibnamefont {Schwerdtfeger}},\ }\href
  {https://doi.org/10.1103/PhysRevLett.122.160801} {\bibfield  {journal}
  {\bibinfo  {journal} {Phys. Rev. Lett.}\ }\textbf {\bibinfo {volume} {122}},\
  \bibinfo {pages} {160801} (\bibinfo {year} {2019})}\BibitemShut {NoStop}%
\bibitem [{\citenamefont {Flambaum}\ and\ \citenamefont
  {Munro-Laylim}(2023)}]{csquarks}%
  \BibitemOpen
  \bibfield  {author} {\bibinfo {author} {\bibfnamefont {V.~V.}\ \bibnamefont
  {Flambaum}}\ and\ \bibinfo {author} {\bibfnamefont {P.}~\bibnamefont
  {Munro-Laylim}},\ }\href {https://doi.org/10.1103/PhysRevD.107.015004}
  {\bibfield  {journal} {\bibinfo  {journal} {Phys. Rev. D}\ }\textbf {\bibinfo
  {volume} {107}},\ \bibinfo {pages} {015004} (\bibinfo {year}
  {2023})}\BibitemShut {NoStop}%
\bibitem [{\citenamefont {Bouley}\ \emph {et~al.}(2023)\citenamefont {Bouley},
  \citenamefont {S\o{}rensen},\ and\ \citenamefont {Yu}}]{BBN-new}%
  \BibitemOpen
  \bibfield  {author} {\bibinfo {author} {\bibfnamefont {T.}~\bibnamefont
  {Bouley}}, \bibinfo {author} {\bibfnamefont {P.}~\bibnamefont
  {S\o{}rensen}},\ and\ \bibinfo {author} {\bibfnamefont {T.-T.}\ \bibnamefont
  {Yu}},\ }\href {https://doi.org/10.1007/JHEP03(2023)104} {\bibfield
  {journal} {\bibinfo  {journal} {JHEP}\ }\textbf {\bibinfo {volume} {03}},\
  \bibinfo {pages} {104}},\ \Eprint {https://arxiv.org/abs/2211.09826}
  {arXiv:2211.09826 [hep-ph]} \BibitemShut {NoStop}%
\bibitem [{\citenamefont {Masia-Roig}\ \emph {et~al.}(2023)\citenamefont
  {Masia-Roig} \emph {et~al.}}]{GNOMEdE}%
  \BibitemOpen
  \bibfield  {author} {\bibinfo {author} {\bibfnamefont {H.}~\bibnamefont
  {Masia-Roig}} \emph {et~al.},\ }\href
  {https://doi.org/10.1103/PhysRevD.108.015003} {\bibfield  {journal} {\bibinfo
   {journal} {Phys. Rev. D}\ }\textbf {\bibinfo {volume} {108}},\ \bibinfo
  {pages} {015003} (\bibinfo {year} {2023})}\BibitemShut {NoStop}%
\bibitem [{\citenamefont {Derevianko}(2018)}]{Devevianko2018}%
  \BibitemOpen
  \bibfield  {author} {\bibinfo {author} {\bibfnamefont {A.}~\bibnamefont
  {Derevianko}},\ }\href {https://doi.org/10.1103/PhysRevA.97.042506}
  {\bibfield  {journal} {\bibinfo  {journal} {Phys. Rev. A}\ }\textbf {\bibinfo
  {volume} {97}},\ \bibinfo {pages} {042506} (\bibinfo {year}
  {2018})}\BibitemShut {NoStop}%
\bibitem [{\citenamefont {Kim}\ and\ \citenamefont {Perez}(2022)}]{KimPerez}%
  \BibitemOpen
  \bibfield  {author} {\bibinfo {author} {\bibfnamefont {H.}~\bibnamefont
  {Kim}}\ and\ \bibinfo {author} {\bibfnamefont {G.}~\bibnamefont {Perez}},\
  }\href@noop {} {\bibinfo {title} {{Oscillations of atomic energy levels
  induced by QCD axion dark matter}}} (\bibinfo {year} {2022}),\ \Eprint
  {https://arxiv.org/abs/2205.12988} {arXiv:2205.12988 [hep-ph]} \BibitemShut
  {NoStop}%
\bibitem [{\citenamefont {Ubaldi}(2010)}]{Ubaldi}%
  \BibitemOpen
  \bibfield  {author} {\bibinfo {author} {\bibfnamefont {L.}~\bibnamefont
  {Ubaldi}},\ }\href {https://doi.org/10.1103/PhysRevD.81.025011} {\bibfield
  {journal} {\bibinfo  {journal} {Phys. Rev. D}\ }\textbf {\bibinfo {volume}
  {81}},\ \bibinfo {pages} {025011} (\bibinfo {year} {2010})}\BibitemShut
  {NoStop}%
\bibitem [{\citenamefont {Flambaum}\ \emph {et~al.}(2004)\citenamefont
  {Flambaum}, \citenamefont {Leinweber}, \citenamefont {Thomas},\ and\
  \citenamefont {Young}}]{Thomas}%
  \BibitemOpen
  \bibfield  {author} {\bibinfo {author} {\bibfnamefont {V.~V.}\ \bibnamefont
  {Flambaum}}, \bibinfo {author} {\bibfnamefont {D.~B.}\ \bibnamefont
  {Leinweber}}, \bibinfo {author} {\bibfnamefont {A.~W.}\ \bibnamefont
  {Thomas}},\ and\ \bibinfo {author} {\bibfnamefont {R.~D.}\ \bibnamefont
  {Young}},\ }\href {https://doi.org/10.1103/PhysRevD.69.115006} {\bibfield
  {journal} {\bibinfo  {journal} {Phys. Rev. D}\ }\textbf {\bibinfo {volume}
  {69}},\ \bibinfo {pages} {115006} (\bibinfo {year} {2004})}\BibitemShut
  {NoStop}%
\bibitem [{\citenamefont {Flambaum}\ and\ \citenamefont
  {Wiringa}(2007)}]{Wiringa1}%
  \BibitemOpen
  \bibfield  {author} {\bibinfo {author} {\bibfnamefont {V.~V.}\ \bibnamefont
  {Flambaum}}\ and\ \bibinfo {author} {\bibfnamefont {R.~B.}\ \bibnamefont
  {Wiringa}},\ }\href {https://doi.org/10.1103/PhysRevC.76.054002} {\bibfield
  {journal} {\bibinfo  {journal} {Phys. Rev. C}\ }\textbf {\bibinfo {volume}
  {76}},\ \bibinfo {pages} {054002} (\bibinfo {year} {2007})}\BibitemShut
  {NoStop}%
\bibitem [{\citenamefont {Flambaum}\ and\ \citenamefont
  {Wiringa}(2009)}]{Wiringa2}%
  \BibitemOpen
  \bibfield  {author} {\bibinfo {author} {\bibfnamefont {V.~V.}\ \bibnamefont
  {Flambaum}}\ and\ \bibinfo {author} {\bibfnamefont {R.~B.}\ \bibnamefont
  {Wiringa}},\ }\href {https://doi.org/10.1103/PhysRevC.79.034302} {\bibfield
  {journal} {\bibinfo  {journal} {Phys. Rev. C}\ }\textbf {\bibinfo {volume}
  {79}},\ \bibinfo {pages} {034302} (\bibinfo {year} {2009})}\BibitemShut
  {NoStop}%
\bibitem [{\citenamefont {Dinh}\ \emph {et~al.}(2009)\citenamefont {Dinh},
  \citenamefont {Dunning}, \citenamefont {Dzuba},\ and\ \citenamefont
  {Flambaum}}]{Dinh}%
  \BibitemOpen
  \bibfield  {author} {\bibinfo {author} {\bibfnamefont {T.~H.}\ \bibnamefont
  {Dinh}}, \bibinfo {author} {\bibfnamefont {A.}~\bibnamefont {Dunning}},
  \bibinfo {author} {\bibfnamefont {V.~A.}\ \bibnamefont {Dzuba}},\ and\
  \bibinfo {author} {\bibfnamefont {V.~V.}\ \bibnamefont {Flambaum}},\ }\href
  {https://doi.org/10.1103/PhysRevA.79.054102} {\bibfield  {journal} {\bibinfo
  {journal} {Phys. Rev. A}\ }\textbf {\bibinfo {volume} {79}},\ \bibinfo
  {pages} {054102} (\bibinfo {year} {2009})}\BibitemShut {NoStop}%
\bibitem [{\citenamefont {Centers}\ \emph {et~al.}(2021)\citenamefont {Centers}
  \emph {et~al.}}]{Derevianko}%
  \BibitemOpen
  \bibfield  {author} {\bibinfo {author} {\bibfnamefont {G.~P.}\ \bibnamefont
  {Centers}} \emph {et~al.},\ }\href
  {https://doi.org/10.1038/s41467-021-27632-7} {\bibfield  {journal} {\bibinfo
  {journal} {Nature Commun.}\ }\textbf {\bibinfo {volume} {12}},\ \bibinfo
  {pages} {7321} (\bibinfo {year} {2021})}\BibitemShut {NoStop}%
\bibitem [{\citenamefont {Diemand}\ \emph {et~al.}(2008)\citenamefont
  {Diemand}, \citenamefont {Kuhlen}, \citenamefont {Madau}, \citenamefont
  {Zemp}, \citenamefont {Moore}, \citenamefont {Potter},\ and\ \citenamefont
  {Stadel}}]{streams}%
  \BibitemOpen
  \bibfield  {author} {\bibinfo {author} {\bibfnamefont {J.}~\bibnamefont
  {Diemand}}, \bibinfo {author} {\bibfnamefont {M.}~\bibnamefont {Kuhlen}},
  \bibinfo {author} {\bibfnamefont {P.}~\bibnamefont {Madau}}, \bibinfo
  {author} {\bibfnamefont {M.}~\bibnamefont {Zemp}}, \bibinfo {author}
  {\bibfnamefont {B.}~\bibnamefont {Moore}}, \bibinfo {author} {\bibfnamefont
  {D.}~\bibnamefont {Potter}},\ and\ \bibinfo {author} {\bibfnamefont
  {J.}~\bibnamefont {Stadel}},\ }\href {https://doi.org/10.1038/nature07153}
  {\bibfield  {journal} {\bibinfo  {journal} {Nature}\ }\textbf {\bibinfo
  {volume} {454}},\ \bibinfo {pages} {735} (\bibinfo {year}
  {2008})}\BibitemShut {NoStop}%
\bibitem [{\citenamefont {Eby}\ \emph {et~al.}(2016)\citenamefont {Eby},
  \citenamefont {Kouvaris}, \citenamefont {Nielsen},\ and\ \citenamefont
  {Wijewardhana}}]{BosonStars}%
  \BibitemOpen
  \bibfield  {author} {\bibinfo {author} {\bibfnamefont {J.}~\bibnamefont
  {Eby}}, \bibinfo {author} {\bibfnamefont {C.}~\bibnamefont {Kouvaris}},
  \bibinfo {author} {\bibfnamefont {N.~G.}\ \bibnamefont {Nielsen}},\ and\
  \bibinfo {author} {\bibfnamefont {L.~C.~R.}\ \bibnamefont {Wijewardhana}},\
  }\href {https://doi.org/10.1007/JHEP02(2016)028} {\bibfield  {journal}
  {\bibinfo  {journal} {JHEP}\ }\textbf {\bibinfo {volume} {02}},\ \bibinfo
  {pages} {028}},\ \Eprint {https://arxiv.org/abs/1511.04474} {arXiv:1511.04474
  [hep-ph]} \BibitemShut {NoStop}%
\bibitem [{\citenamefont {Pospelov}\ \emph {et~al.}(2013)\citenamefont
  {Pospelov}, \citenamefont {Pustelny}, \citenamefont {Ledbetter},
  \citenamefont {Jackson~Kimball}, \citenamefont {Gawlik},\ and\ \citenamefont
  {Budker}}]{TopDefects}%
  \BibitemOpen
  \bibfield  {author} {\bibinfo {author} {\bibfnamefont {M.}~\bibnamefont
  {Pospelov}}, \bibinfo {author} {\bibfnamefont {S.}~\bibnamefont {Pustelny}},
  \bibinfo {author} {\bibfnamefont {M.~P.}\ \bibnamefont {Ledbetter}}, \bibinfo
  {author} {\bibfnamefont {D.~F.}\ \bibnamefont {Jackson~Kimball}}, \bibinfo
  {author} {\bibfnamefont {W.}~\bibnamefont {Gawlik}},\ and\ \bibinfo {author}
  {\bibfnamefont {D.}~\bibnamefont {Budker}},\ }\href
  {https://doi.org/10.1103/PhysRevLett.110.021803} {\bibfield  {journal}
  {\bibinfo  {journal} {Phys. Rev. Lett.}\ }\textbf {\bibinfo {volume} {110}},\
  \bibinfo {pages} {021803} (\bibinfo {year} {2013})}\BibitemShut {NoStop}%
\bibitem [{\citenamefont {Oswald}\ \emph {et~al.}(2022)\citenamefont {Oswald}
  \emph {et~al.}}]{I2}%
  \BibitemOpen
  \bibfield  {author} {\bibinfo {author} {\bibfnamefont {R.}~\bibnamefont
  {Oswald}} \emph {et~al.},\ }\href
  {https://doi.org/10.1103/PhysRevLett.129.031302} {\bibfield  {journal}
  {\bibinfo  {journal} {Phys. Rev. Lett.}\ }\textbf {\bibinfo {volume} {129}},\
  \bibinfo {pages} {031302} (\bibinfo {year} {2022})}\BibitemShut {NoStop}%
\bibitem [{\citenamefont {Vermeulen}\ \emph {et~al.}(2021)\citenamefont
  {Vermeulen} \emph {et~al.}}]{GEO600}%
  \BibitemOpen
  \bibfield  {author} {\bibinfo {author} {\bibfnamefont {S.}~\bibnamefont
  {Vermeulen}} \emph {et~al.},\ }\href
  {https://doi.org/https://doi.org/10.1038/s41586-021-04031-y} {\bibfield
  {journal} {\bibinfo  {journal} {Nature}\ }\textbf {\bibinfo {volume} {600}},\
  \bibinfo {pages} {424–428} (\bibinfo {year} {2021})}\BibitemShut {NoStop}%
\bibitem [{\citenamefont {Abel}\ \emph {et~al.}(2017)\citenamefont {Abel} \emph
  {et~al.}}]{nEDM}%
  \BibitemOpen
  \bibfield  {author} {\bibinfo {author} {\bibfnamefont {C.}~\bibnamefont
  {Abel}} \emph {et~al.},\ }\href {https://doi.org/10.1103/PhysRevX.7.041034}
  {\bibfield  {journal} {\bibinfo  {journal} {Phys. Rev. X}\ }\textbf {\bibinfo
  {volume} {7}},\ \bibinfo {pages} {041034} (\bibinfo {year}
  {2017})}\BibitemShut {NoStop}%
\bibitem [{\citenamefont {Blum}\ \emph {et~al.}(2014)\citenamefont {Blum},
  \citenamefont {D'Agnolo}, \citenamefont {Lisanti},\ and\ \citenamefont
  {Safdi}}]{BBN}%
  \BibitemOpen
  \bibfield  {author} {\bibinfo {author} {\bibfnamefont {K.}~\bibnamefont
  {Blum}}, \bibinfo {author} {\bibfnamefont {R.~T.}\ \bibnamefont {D'Agnolo}},
  \bibinfo {author} {\bibfnamefont {M.}~\bibnamefont {Lisanti}},\ and\ \bibinfo
  {author} {\bibfnamefont {B.~R.}\ \bibnamefont {Safdi}},\ }\href
  {https://doi.org/https://doi.org/10.1016/j.physletb.2014.07.059} {\bibfield
  {journal} {\bibinfo  {journal} {Phys. Lett. B}\ }\textbf {\bibinfo {volume}
  {737}},\ \bibinfo {pages} {30} (\bibinfo {year} {2014})}\BibitemShut
  {NoStop}%
\bibitem [{\citenamefont {Lucente}\ \emph {et~al.}(2022)\citenamefont
  {Lucente}, \citenamefont {Mastrototaro}, \citenamefont {Carenza},
  \citenamefont {Di~Luzio}, \citenamefont {Giannotti},\ and\ \citenamefont
  {Mirizzi}}]{SN1987}%
  \BibitemOpen
  \bibfield  {author} {\bibinfo {author} {\bibfnamefont {G.}~\bibnamefont
  {Lucente}}, \bibinfo {author} {\bibfnamefont {L.}~\bibnamefont
  {Mastrototaro}}, \bibinfo {author} {\bibfnamefont {P.}~\bibnamefont
  {Carenza}}, \bibinfo {author} {\bibfnamefont {L.}~\bibnamefont {Di~Luzio}},
  \bibinfo {author} {\bibfnamefont {M.}~\bibnamefont {Giannotti}},\ and\
  \bibinfo {author} {\bibfnamefont {A.}~\bibnamefont {Mirizzi}},\ }\href
  {https://doi.org/10.1103/PhysRevD.105.123020} {\bibfield  {journal} {\bibinfo
   {journal} {Phys. Rev. D}\ }\textbf {\bibinfo {volume} {105}},\ \bibinfo
  {pages} {123020} (\bibinfo {year} {2022})}\BibitemShut {NoStop}%
\bibitem [{\citenamefont {Mehta}\ \emph {et~al.}(2020)\citenamefont {Mehta},
  \citenamefont {Demirtas}, \citenamefont {Long}, \citenamefont {Marsh},
  \citenamefont {Mcallister},\ and\ \citenamefont {Stott}}]{spins1}%
  \BibitemOpen
  \bibfield  {author} {\bibinfo {author} {\bibfnamefont {V.~M.}\ \bibnamefont
  {Mehta}}, \bibinfo {author} {\bibfnamefont {M.}~\bibnamefont {Demirtas}},
  \bibinfo {author} {\bibfnamefont {C.}~\bibnamefont {Long}}, \bibinfo {author}
  {\bibfnamefont {D.~J.~E.}\ \bibnamefont {Marsh}}, \bibinfo {author}
  {\bibfnamefont {L.}~\bibnamefont {Mcallister}},\ and\ \bibinfo {author}
  {\bibfnamefont {M.~J.}\ \bibnamefont {Stott}},\ }\href@noop {} {\bibinfo
  {title} {Superradiance exclusions in the landscape of type {IIB} string
  theory}} (\bibinfo {year} {2020}),\ \Eprint
  {https://arxiv.org/abs/2011.08693} {arXiv:2011.08693 [hep-th]} \BibitemShut
  {NoStop}%
\bibitem [{\citenamefont {Baryakhtar}\ \emph {et~al.}(2021)\citenamefont
  {Baryakhtar}, \citenamefont {Galanis}, \citenamefont {Lasenby},\ and\
  \citenamefont {Simon}}]{spins2}%
  \BibitemOpen
  \bibfield  {author} {\bibinfo {author} {\bibfnamefont {M.}~\bibnamefont
  {Baryakhtar}}, \bibinfo {author} {\bibfnamefont {M.}~\bibnamefont {Galanis}},
  \bibinfo {author} {\bibfnamefont {R.}~\bibnamefont {Lasenby}},\ and\ \bibinfo
  {author} {\bibfnamefont {O.}~\bibnamefont {Simon}},\ }\href
  {https://doi.org/10.1103/PhysRevD.103.095019} {\bibfield  {journal} {\bibinfo
   {journal} {Phys. Rev. D}\ }\textbf {\bibinfo {volume} {103}},\ \bibinfo
  {pages} {095019} (\bibinfo {year} {2021})}\BibitemShut {NoStop}%
\bibitem [{\citenamefont {Hook}\ and\ \citenamefont {Huang}(2018)}]{Earth}%
  \BibitemOpen
  \bibfield  {author} {\bibinfo {author} {\bibfnamefont {A.}~\bibnamefont
  {Hook}}\ and\ \bibinfo {author} {\bibfnamefont {J.}~\bibnamefont {Huang}},\
  }\href {https://doi.org/10.1007/JHEP06(2018)036} {\bibfield  {journal}
  {\bibinfo  {journal} {JHEP}\ }\textbf {\bibinfo {volume} {06}},\ \bibinfo
  {pages} {036}},\ \Eprint {https://arxiv.org/abs/1708.08464} {arXiv:1708.08464
  [hep-ph]} \BibitemShut {NoStop}%
\bibitem [{\citenamefont {Foster}\ \emph {et~al.}(2018)\citenamefont {Foster},
  \citenamefont {Rodd},\ and\ \citenamefont {Safdi}}]{Foster}%
  \BibitemOpen
  \bibfield  {author} {\bibinfo {author} {\bibfnamefont {J.~W.}\ \bibnamefont
  {Foster}}, \bibinfo {author} {\bibfnamefont {N.~L.}\ \bibnamefont {Rodd}},\
  and\ \bibinfo {author} {\bibfnamefont {B.~R.}\ \bibnamefont {Safdi}},\ }\href
  {https://doi.org/10.1103/PhysRevD.97.123006} {\bibfield  {journal} {\bibinfo
  {journal} {Phys. Rev. D}\ }\textbf {\bibinfo {volume} {97}},\ \bibinfo
  {pages} {123006} (\bibinfo {year} {2018})}\BibitemShut {NoStop}%
\bibitem [{\citenamefont {Crewther}\ \emph {et~al.}(1979)\citenamefont
  {Crewther}, \citenamefont {{Di Vecchia}}, \citenamefont {Veneziano},\ and\
  \citenamefont {Witten}}]{Witten}%
  \BibitemOpen
  \bibfield  {author} {\bibinfo {author} {\bibfnamefont {R.}~\bibnamefont
  {Crewther}}, \bibinfo {author} {\bibfnamefont {P.}~\bibnamefont {{Di
  Vecchia}}}, \bibinfo {author} {\bibfnamefont {G.}~\bibnamefont {Veneziano}},\
  and\ \bibinfo {author} {\bibfnamefont {E.}~\bibnamefont {Witten}},\ }\href
  {https://doi.org/https://doi.org/10.1016/0370-2693(79)90128-X} {\bibfield
  {journal} {\bibinfo  {journal} {Phys. Lett. B}\ }\textbf {\bibinfo {volume}
  {88}},\ \bibinfo {pages} {123} (\bibinfo {year} {1979})}\BibitemShut
  {NoStop}%
\bibitem [{\citenamefont {Seiferle}\ \emph {et~al.}(2019)\citenamefont
  {Seiferle} \emph {et~al.}}]{Seiferle2019}%
  \BibitemOpen
  \bibfield  {author} {\bibinfo {author} {\bibfnamefont {B.}~\bibnamefont
  {Seiferle}} \emph {et~al.},\ }\href
  {https://doi.org/10.1038/s41586-019-1533-4} {\bibfield  {journal} {\bibinfo
  {journal} {Nature}\ }\textbf {\bibinfo {volume} {573}},\ \bibinfo {pages}
  {243} (\bibinfo {year} {2019})}\BibitemShut {NoStop}%
\bibitem [{\citenamefont {Flambaum}\ \emph {et~al.}(1984)\citenamefont
  {Flambaum}, \citenamefont {Khriplovich},\ and\ \citenamefont
  {Sushkov}}]{JETP1984}%
  \BibitemOpen
  \bibfield  {author} {\bibinfo {author} {\bibfnamefont {V.~V.}\ \bibnamefont
  {Flambaum}}, \bibinfo {author} {\bibfnamefont {I.~B.}\ \bibnamefont
  {Khriplovich}},\ and\ \bibinfo {author} {\bibfnamefont {O.~P.}\ \bibnamefont
  {Sushkov}},\ }\href@noop {} {\bibfield  {journal} {\bibinfo  {journal} {Sov.
  Phys. JETP}\ }\textbf {\bibinfo {volume} {60}},\ \bibinfo {pages} {873}
  (\bibinfo {year} {1984})}\BibitemShut {NoStop}%
\bibitem [{\citenamefont {Pospelov}\ and\ \citenamefont
  {Ritz}(2005)}]{PospelovRitz}%
  \BibitemOpen
  \bibfield  {author} {\bibinfo {author} {\bibfnamefont {M.}~\bibnamefont
  {Pospelov}}\ and\ \bibinfo {author} {\bibfnamefont {A.}~\bibnamefont
  {Ritz}},\ }\href {https://doi.org/10.1016/j.aop.2005.04.002} {\bibfield
  {journal} {\bibinfo  {journal} {Annals Phys.}\ }\textbf {\bibinfo {volume}
  {318}},\ \bibinfo {pages} {119} (\bibinfo {year} {2005})}\BibitemShut
  {NoStop}%
\bibitem [{\citenamefont {Flambaum}\ \emph {et~al.}(2014)\citenamefont
  {Flambaum}, \citenamefont {DeMille},\ and\ \citenamefont {Kozlov}}]{DeMille}%
  \BibitemOpen
  \bibfield  {author} {\bibinfo {author} {\bibfnamefont {V.~V.}\ \bibnamefont
  {Flambaum}}, \bibinfo {author} {\bibfnamefont {D.}~\bibnamefont {DeMille}},\
  and\ \bibinfo {author} {\bibfnamefont {M.~G.}\ \bibnamefont {Kozlov}},\
  }\href {https://doi.org/10.1103/PhysRevLett.113.103003} {\bibfield  {journal}
  {\bibinfo  {journal} {Phys. Rev. Lett.}\ }\textbf {\bibinfo {volume} {113}},\
  \bibinfo {pages} {103003} (\bibinfo {year} {2014})}\BibitemShut {NoStop}%
\bibitem [{\citenamefont {de~Vries}\ \emph {et~al.}(2015)\citenamefont
  {de~Vries}, \citenamefont {Mereghetti},\ and\ \citenamefont
  {Walker-Loud}}]{deVries}%
  \BibitemOpen
  \bibfield  {author} {\bibinfo {author} {\bibfnamefont {J.}~\bibnamefont
  {de~Vries}}, \bibinfo {author} {\bibfnamefont {E.}~\bibnamefont
  {Mereghetti}},\ and\ \bibinfo {author} {\bibfnamefont {A.}~\bibnamefont
  {Walker-Loud}},\ }\href {https://doi.org/10.1103/PhysRevC.92.045201}
  {\bibfield  {journal} {\bibinfo  {journal} {Phys. Rev. C}\ }\textbf {\bibinfo
  {volume} {92}},\ \bibinfo {pages} {045201} (\bibinfo {year}
  {2015})}\BibitemShut {NoStop}%
\bibitem [{\citenamefont {Angeli}\ and\ \citenamefont
  {Marinova}(2013)}]{NuclearRadius}%
  \BibitemOpen
  \bibfield  {author} {\bibinfo {author} {\bibfnamefont {I.}~\bibnamefont
  {Angeli}}\ and\ \bibinfo {author} {\bibfnamefont {K.}~\bibnamefont
  {Marinova}},\ }\href
  {https://doi.org/https://doi.org/10.1016/j.adt.2011.12.006} {\bibfield
  {journal} {\bibinfo  {journal} {At. Data Nucl. Data Tables}\ }\textbf
  {\bibinfo {volume} {99}},\ \bibinfo {pages} {69} (\bibinfo {year}
  {2013})}\BibitemShut {NoStop}%
\end{thebibliography}
\end{document}